\documentclass[preprint,authoryear,12pt]{elsarticle}

\usepackage[margin=1in]{geometry}
\usepackage{amsmath,amssymb,amsfonts,bm}
\usepackage{booktabs}
\usepackage{graphicx}
\usepackage{float}
\usepackage{hyperref}
\usepackage{xcolor}
\usepackage{enumitem}
\usepackage{siunitx}
\usepackage{comment}
\usepackage{lineno}

\hypersetup{colorlinks=true,linkcolor=blue,citecolor=blue,urlcolor=blue}

\newcommand{\eps}{\boldsymbol{\varepsilon}}
\newcommand{\sig}{\boldsymbol{\sigma}}
\newcommand{\sdev}{\boldsymbol{s}}
\newcommand{\z}{\boldsymbol{z}}
\newcommand{\x}{\boldsymbol{x}}

\newcommand{\I}{\mathbf{I}}
\newcommand{\eeq}{\varepsilon^{\mathrm{eq}}}
\newcommand{\eeqmax}{\varepsilon^{\mathrm{eq,max}}}
\newcommand{\eeqt}{\tilde{\varepsilon}^{\mathrm{eq}}}
\newcommand{\tr}{\operatorname{tr}}

\begin{document}

\begin{frontmatter}

\title{A Gaussian Process State Space Formulation for Materials Constitutive Modeling: Application to Quasi-brittle Materials}

\author[1]{Houlin Xu\corref{cor1}}
\ead{hxu122@jh.edu}
\author[1]{Chenyang Li}
\author[1,2]{K.T. Ramesh}
\author[1,3]{Michael D. Shields}
\cortext[cor1]{Corresponding author}

\affiliation[1]{organization={Hopkins Extreme Materials Institute, Johns Hopkins University},
                city={Baltimore}, state={MD}, country={USA}}
\affiliation[2]{organization={Department of Mechanical Engineering, Johns Hopkins University},
                city={Baltimore}, state={MD}, country={USA}}
\affiliation[3]{organization={Department of Civil and Systems Engineering, Johns Hopkins University},
                city={Baltimore}, state={MD}, country={USA}}

\begin{abstract}
Constitutive models for history-dependent, path-dependent materials typically encode this dependence in prescribed functional forms through internal variables such as damage or plastic strain that are phenomenological, model-dependent, and not directly observable. These rigid functional forms lack flexibility to match observations, make calibration potentially non-unique, and complicate uncertainty quantification and propagation through finite element predictions. This paper develops a Gaussian-process state-space model (GP-SSM) that replaces the fixed form of the constitutive model with a flexible GP regression, and the prescribed internal variables with a set of latent state variables learned directly from observable stress--strain histories. As a GP, the model is naturally endowed with a measure of uncertainty and can be constrained to satisfy necessary physical principles. We demonstrate this framework for quasi-brittle materials under triaxial compression, where the latent variable evolution is driven by the current-state stress invariants and the increment of equivalent strain 
to reflect the evolution of irreversible, pressure-sensitive history. A sparse multi-output Gaussian process then maps the latent state and current strain tensor to the six-component stress response. Variational sparse inference enables tractable training from sequential stress-strain data that retains a probabilistic structure for offline uncertainty assessment.
The trained model is exported into Abaqus/Explicit through a user material subroutine. 
Four verification cases are reported: (1) offline full-sequence rollout evaluation on the validation and test histories; (2) a single element simulation under changing confinement, which exercises the complete Abaqus/Explicit deployment pathway; (3) a frozen-memory unloading--reloading extension in which latent evolution stops once equivalent strain recedes below its historical maximum; and (4) a multi-element structural model that reproduces a loading condition from the Karagozian and Case Concrete (KCC) model and is compared directly against KCC as a reference solution. 
These results show that the learned state-space update (i) reproduces held-out RVE-scale histories in closed-loop rollout, (ii) can be evaluated incrementally in an explicit finite element solver, (iii) tracks a confining stress that evolves during loading, (iv) supports unloading without violating the irreversibility of the learned history state, and (v) reproduces structural-level compression behavior consistent with an established constitutive model. The framework provides a practical route to bring together probabilistic machine learning, latent internal-state modeling, and finite element-compatible constitutive modeling.

\end{abstract}


\end{frontmatter}

\section{Introduction}

Path-dependent, history-dependent constitutive behavior is difficult to model because the current stress depends not only on the current strain but on the loading path. Classical phenomenological models, including damage mechanics, plasticity, and cap-type formulations, have hypothesized functional forms that represent this path dependence through internal variables whose predictive capability depends strongly on parameters that are difficult to identify uniquely from limited tests. This difficulty is especially pronounced for quasi-brittle materials such as concrete and rock \citep{bazant1998fracture, yang2025mechanism}, where the macroscopic response is governed by evolving microcracking, frictional sliding, compaction, dilatancy, and confinement-dependent failure. These mechanisms produce nonlinear stiffness degradation, pressure-sensitive strength, path dependence, and post-peak softening. Cap-type formulations such as the Karagozian and Case Concrete (KCC) model \citep{malvar1997kcc,wu2015numerical} have been widely adopted for this material class because they encode mechanics-inspired assumptions and are computationally efficient in finite element solvers, although (like other phenomenological models) they inherit the same difficulty of calibration.

The central difficulty is not only that material parameters are uncertain. The internal history variables used by many constitutive theories are not directly measurable in standard stress--strain experiments. Quantities such as damage, plastic strain, hardening, crack-density, or compaction are introduced to make the constitutive update path dependent, but despite their physical interpretability, these variables as typically defined are model-specific and difficult to measure.  As a result, calibration often becomes an ill-posed inverse problem: different parameter sets and internal-variable evolutions may reproduce the same observed stress--strain curves while implying different predictions under other loading paths.

Replacing such fixed-form models and their prescribed internal variables with a representation learned directly from stress--strain data has a substantial history, dating back to early neural-network constitutive models \citep{ghaboussi1991knowledge}. Deep-learning surrogates have also been applied directly to confinement-dependent, strain-softening rock behavior under triaxial loading \citep{shi2022deeplearning}. Recent surveys catalog the  breadth of data-driven constitutive laws across material classes \citep{fuhg2025review}. A comprehensive review of these models is beyond the scope of this work.

The need for a history-dependent representation of the material state
is particularly important for quasi-brittle materials under triaxial compression. Confinement can change both stiffness and strength, while deviatoric loading controls shear-dominated damage and post-peak evolution. Two material points with the same instantaneous strain tensor may therefore correspond to different crack, compaction, and stress-history states. A data-driven material law that predicts stress only from the current strain tensor is unlikely to be robust when embedded in a finite element analysis, where local loading paths are generated by the coupled boundary-value problem rather than prescribed directly from laboratory tests.

State-space learning provides a natural way to address this problem. Since the classical Kalman filter \citep{kalman1960}, state-space models have introduced a hidden state that evolves incrementally and summarizes the information needed to predict a system's future observations; Gaussian-process (GP) and recurrent neural-network extensions of this idea have been used to learn nonlinear latent dynamics directly from sequential data \citep{wang2008gpdm,krishnan2017structured}. In a constitutive setting, such a hidden state can be interpreted as a learned low-dimensional representation of the material history. Note that the hidden state need not be identifiable as the damage, plastic strain, or any particular classical or hypothesized internal variable. However, it serves the same functional role, encoding the essential information from the past that is needed to update the current stress. Moreover, essential physical constraints (e.g. non-negative dissipation) can be placed on these latent variables, and their interpretation may provide important mechanistic insights into material performance.

In a related effort, recent work has begun to replace individual phenomenological surfaces within classical concrete models with data-driven models. \citet{li2026physics} replace the empirical failure surface of the KCC model with a physics-constrained GP regression surrogate trained on triaxial compression data, and show that embedding known material-property constraints into the GP prior markedly improves accuracy (compared with an unconstrained regressor) and reduces predictive uncertainty and inadmissible material behavior beyond the training confinement range. That work uses a surrogate that replaces a single static failure surface evaluated at a given loading state.

This paper develops a Gaussian-process state-space model (GP-SSM) formulation for history-dependent constitutive modeling. The model is specifically applied in the context of quasi-brittle materials, but the framework can be applied broadly to diverse classes of nonlinear history-dependent materials. The model combines three ingredients. First, a low-dimensional latent state represents unobserved material history. In conventional
recurrent neural-network surrogates for path-dependent plasticity, the hidden vector plays two
roles simultaneously: it propagates the material history forward from one increment to the next,
and its size also determines how complex an update the network can represent.
A more capable surrogate therefore requires a larger stored state, and such models
commonly use comparatively large hidden states \citep{mozaffar2019deep,gorji2020rnn}. 
The
Minimal State Cell of \citet{bonatti2021oneforall} was designed to separate these
two roles, so that a wide and deep update network can act on a persistent state of only a few
variables.

Our model shares this state-parsimony objective. Its persistent memory is restricted to a small, fixed number of latent variables learned directly from stress--strain histories rather than prescribed as phenomenological internal variables. Its distinction lies in combining this compact state with a mechanics-informed evolution.

The latent variable evolution is driven by equivalent strain and previous-step stress invariants, introducing mechanics-informed sensitivity to irreversible loading progression, pressure, and deviatoric stress intensity. Then, a sparse multi-output Gaussian process \citep{rasmussen2005gpml,alvarez2012kernels}, trained via stochastic variational inference over inducing points \citep{titsias2009variational,hensman2013big}, maps the latent state and current strain tensor to the six-component stress tensor. This is in contrast to deterministic neural-network stress maps, and follows the general precedent of using Gaussian processes to model latent dynamical systems \citep{deisenroth2015gpcontrol}. The Gaussian-process structure provides a nonparametric and data-efficient stress map built from a probabilistic formulation for quantifying uncertainty in the material model and allows offline uncertainty propagation through finite element simulations.

A practical objective of this work is finite element compatibility. Data-driven constitutive models are frequently developed and validated as offline regressors. History-dependent machine-learning surrogates that have been embedded directly in finite element solvers have primarily been applied to accelerating multiscale FE\textsuperscript{2} homogenization by, e.g., replacing an expensive micro-scale solve with a recurrent-neural-network surrogate \citep{ghavamian2019accelerating}, or by using a trained recurrent network as the homogenized material update itself \citep{logarzo2021smart}. This work demonstrates a full experiment-to-deployment pipeline built as an alternative to phenomenological constitutive models, in which the material model is evaluated locally and incrementally at every integration point, with state variables stored and updated by the solver. Note that although the verification cases in this paper use concrete-like triaxial data and are compared against the KCC model, the mechanics-informed state-space structure of GP-SSM is applicable to other pressure-sensitive, path-dependent quasi-brittle materials such as rocks and ceramics, that share the same broad failure mechanisms \citep{bazant1998fracture, li2021finite}.

The remainder of the paper is organized as follows. Section~\ref{sec:background} reviews the elastic-predictor, boundary-correction structure shared by classical plasticity and damage-plasticity models, using the KCC model as a representative example, and highlights the phenomenological cost of designing pressure-dependent failure boundaries explicitly. Section~\ref{sec:problem} formalizes the latent-state constitutive problem in this alternative setting. Section~\ref{sec:formulation} presents the GP-SSM formulation. Section~\ref{sec:data} describes data processing and training. Section~\ref{sec:results} reports four verification cases, beginning with offline full-sequence rollout evaluation and proceeding through Abaqus/Explicit material-point and structural simulations. Finally, Sections~\ref{sec:discussion} and~\ref{sec:conclusion} discuss limitations and conclusions.

\section{Background: constitutive modeling paradigms}
\label{sec:background}

\subsection{Mechanistic, phenomenological, and data-driven modeling}
\label{sec:two-axes}

Constructing a constitutive model requires developing the stress--strain relation, and its history dependence. Three broad approaches are used in practice. \emph{Mechanistic} models derive the relation from an explicit subscale mechanism. \emph{Phenomenological} models instead postulate the relation at the scale of application and calibrate it to fit observed response, without deriving it directly from a lower-scale mechanism. \emph{Data-driven} models learn the relation directly from stress--strain data, with minimal or no assumed functional form. These labels classify the origins of a modeling approach rather than the model that contains it. A complete constitutive model is typically a composite: it may derive one component from a mechanism, postulate a second, and (particularly recently) learn a third. The three approaches also differ in how much structure is assumed before any data are seen, and that choice involves a consistent trade-off. The more structure a model assumes, the less data it requires and the further it can be trusted outside the conditions on which it was calibrated, but the more its predictions depend on that structure being correct.

The mechanistic paradigm is broadly applicable, but any individual mechanistic model is necessarily specific to the material class whose deformation mechanism it encodes. In metals, crystal plasticity derives macroscopic flow directly from dislocation slip on crystallographic slip systems \citep{roters2010cpfem,roters2019damask}. In polymers and elastomers, network theories such as the Arruda--Boyce eight-chain model derive the hyperelastic stress--stretch relation from the statistical mechanics of a network of freely jointed polymer chains \citep{arruda1993threedimensional}. In both cases, the relation follows as a consequence of the assumed lower-scale mechanism rather than as a curve fit to test data, and in both cases the resulting model inherits the scope of that mechanism: slip kinematics has no counterpart in an elastomer network, and vice versa. The mechanistic route is therefore available only where the governing mechanism is both identified and tractably represented at the scale of interest.

Phenomenological constitutive models have likewise been developed for a variety of material classes. Mooney--Rivlin hyperelasticity postulates a strain-energy function directly in terms of invariants of the deformation and calibrates its coefficients to (for example) rubber test data \citep{mooney1940theory,rivlin1948large}, without deriving that function from a network mechanism. $J_2$ (von Mises) plasticity with isotropic or kinematic hardening postulates a yield surface and hardening law for metals and calibrates them to observed hardening curves \citep{lubliner2008plasticity,simo1998computational}, without necessarily deriving them from the mechanics of dislocation slip.

The boundary between the mechanistic and phenomenological approaches may be softer than the definitions suggest. A postulated form is rarely arbitrary: e.g., $J_2$ plasticity assumes that yielding is insensitive to pressure and to the third invariant, an assumption that is mechanistic in origin, since dislocation slip is volume-preserving. What makes such a model phenomenological is not the absence of mechanistic reasoning, but that the mechanism informs the choice of form rather than supplying it. This mixed character is apparent, for example, in damage or porosity-based constitutive models such as the Gurson-Tvergaard-Needleman (GTN) model~\citep{gurson1977continuum,tvergaard1981influence,tvergaard1984analysis} that hypothesize a pressure-dependent yield function that is porosity dependent and phenomenologically informed by the mechanisms of void nucleation, growth, and coalescence.  

Data-driven constitutive models learn the relation, or its evolution, directly from observed data (whether measured or generated by simulation) with little or no assumed functional form. Such models differ among themselves chiefly in what is learned. Sequence models learn the incremental update itself, mapping strain history to stress through a recurrent hidden state \citep{ghaboussi1991knowledge,mozaffar2019deep,gorji2020rnn,bonatti2021oneforall}. Potential-based approaches learn a stored-energy or dissipation function and recover stress by differentiation, so that the learned object is a scalar rather than the stress \citep{masi2021thermodynamics,asad2022mechanics}. Discovery-based approaches learn a sparse, interpretable expression for the law itself, in some cases from full-field displacement and reaction data without stress labels \citep{flaschel2022plasticity}. Model-free data-driven computing dispenses with a constitutive function altogether, searching a database of observed states~\citep{kirchdoerfer2016datadriven} or their distribution~\citep{xu2024stochastic} directly; it is less a way of constructing a constitutive relation than a reformulation of the boundary-value problem that avoids needing one. \citet{fuhg2025review} survey this landscape across material classes in far more detail than is attempted here. 

The microplane model M7 \citep{caner2013microplane1,caner2013microplane2} illustrates the  composite character of practical models.
Its kinematic constraint decomposes strain onto a predefined set of microplane orientations and recovers the macroscopic stress through the principle of virtual work. This component is mechanistic, reflecting a geometric description of how the material transmits stress across orientations. The `boundary' functions that bound the admissible normal and shear stresses on each microplane, however, do not follow from the kinematic constraint. Their functional forms and parameters are selected to reproduce observed uniaxial, biaxial, and triaxial responses. In the terminology used here, these functions provide the phenomenological component of the model. The same pattern recurs in the KCC model \citep{malvar1997kcc,malvar2000kcc,wu2012performance,wu2015numerical} and in LDPM \citep{cusatis2011ldpm1,cusatis2011ldpm2}. An explicit structural hypothesis (a pressure-dependent invariant space for KCC, particle geometry and contact kinematics for LDPM) is combined with strength or boundary functions calibrated against experimental data.

Hybridization also occurs on the data-driven side. A learned component can be incorporated into a mechanistic model, for example by replacing the microplane boundary functions with a learned surrogate while retaining the kinematic constraint and virtual-work recovery. Alternatively, a learned model can keep the structural skeleton of a phenomenological model (an internal variable, an evolution law, and a stress response) while declining to specify any of them in closed form. Learned models are also routinely constrained to satisfy general mechanical requirements: a convex learned potential yields a hyperelastic, materially stable stress \citep{asad2022mechanics}; invariant inputs enforce frame indifference and material symmetry \citep{heider2020so3,vlassis2020geometric}; a sign-constrained evolution enforces irreversibility and nonnegative dissipation \citep{masi2021thermodynamics}. Constraints of this kind are not themselves a form of hybridization. Every constitutive model, in all three paradigms, is required to satisfy them. They restrict which relations are admissible, not where a relation originates.

In this work, no boundary function is postulated and no return-mapping correction is performed. The stress update itself is learned as a nonparametric map from the current strain and an internal state to the current stress; whatever admissible region the material obeys is therefore implicit in that learned map under physics constraints.
What the formulation retains from the models above is their structural skeleton rather than a derivation from any particular mechanism. It is mechanics-informed: the descriptors that drive the state are frame-invariant and the history they accumulate is irreversible. Section~\ref{sec:phenomenological-cost} makes this positioning precise.

\subsection{Constitutive models for quasi-brittle materials}
\label{sec:two-axes-specialization}
\label{sec:structure}
Although the method is general, this paper focuses on quasi-brittle materials such as concrete and rock, for which the phenomenological paradigm has long been the dominant approach.
Most established constitutive models for concrete and other quasi-brittle materials  (including pressure-dependent plasticity \citep{drucker1952soil}, cap-plasticity \citep{dimaggio1971granular}, and continuum damage-plasticity formulations \citep{lubliner1989plasticdamage}) share a common predictor-corrector algorithmic structure. 

At each load increment, a trial stress is first computed elastically from the strain increment and the current elastic stiffness as,
\begin{equation}
\sig_t^{\mathrm{trial}}=\sig_{t-1}+\mathbf{C}^{e}:\Delta\eps_t .
\label{eq:elastic_trial}
\end{equation}
The trial stress is then tested against an admissible region, which is defined by a scalar yield or failure function, \(F(\sig,\mathcal{H})\le0\), typically written in terms of stress invariants and a set of internal history variables \(\mathcal{H}\), such as accumulated plastic strain and/or a damage parameter. This function effectively defines a boundary on the elastic domain in stress space. If the trial stress violates the boundary, \(F(\sig_t^{\mathrm{trial}},\mathcal{H}_{t-1})>0\), it is mapped back onto the boundary (\(F(\sig,\mathcal{H})=0\)) through a correction step -- typically a radial return, closest-point projection, or cutting-plane algorithm -- and \(\mathcal{H}\) is updated through an associated or non-associated flow and hardening/damage rule. Because the elastic operator \(\mathbf{C}^e\) is 
prescribed by a small number of elastic constants (even when it is degraded by damage through $\mathcal{H})$
and largely independent of material form, essentially all model-specific physics (and the engineering judgment/inference required to calibrate it) is concentrated in the design and evolution of the boundary function \(F\).

\subsection{A representative example: the KCC model}

The Karagozian \& Case Concrete (KCC) model \citep{malvar1997kcc,malvar2000kcc,wu2012performance,wu2015numerical} is a widely used model for concrete under impact and blast loading that exhibits this phenomenological predictor-corrector structure. 
KCC first fixes three pressure-dependent reference surfaces in deviatoric stress versus pressure space: a yield surface \(\Delta\sigma_y(p)\), a maximum (failure) surface \(\Delta\sigma_m(p)\), and a residual surface \(\Delta\sigma_r(p)\), where \(p\) denotes the mean pressure. Then the current bounding surface is obtained by interpolating between them as damage accumulates. Each reference surface is specified in closed form as a rational function of the pressure,
\begin{equation}
\Delta\sigma_y(p)=a_{0y}+\frac{p}{a_{1y}+a_{2y}p},\qquad
\Delta\sigma_m(p)=a_{0}+\frac{p}{a_{1}+a_{2}p},\qquad
\Delta\sigma_r(p)=\frac{p}{a_{1f}+a_{2f}p},
\label{eq:kcc_surfaces}
\end{equation}
with the constants \(a_{0y},a_{1y},a_{2y},a_0,a_1,a_2,a_{1f},a_{2f}\) calibrated by curve-fitting unconfined and confined compression and tension test data over the pressure range of interest. A scalar damage-like history variable \(\lambda\) accumulates with the effective plastic strain increment, and is scaled differently in tension and in compression.  
A normalized modulation function \(\eta(\lambda)\in[0,1]\), which rises from zero to unity as
\(\lambda\) increases to \(\lambda_m\) and decays back to zero thereafter, interpolates the
current strength surface between the yield and maximum surfaces during hardening
(\(\lambda\le\lambda_m\)) and between the maximum and residual surfaces during softening
(\(\lambda>\lambda_m\)). The transition value \(\lambda_m\) is the point at which the current
surface coincides with the maximum surface, and corresponds to the peak stress of the
stress--strain response, resulting in the pressure and damage dependent surface taking the following form: 
\begin{equation}
\Delta\sigma(p,\lambda)=
\begin{cases}
\Delta\sigma_y(p)+\eta(\lambda)\left[\Delta\sigma_m(p)-\Delta\sigma_y(p)\right], & \lambda\le\lambda_m,\\[4pt]
\Delta\sigma_r(p)+\eta(\lambda)\left[\Delta\sigma_m(p)-\Delta\sigma_r(p)\right], & \lambda>\lambda_m.
\end{cases}
\label{eq:kcc_eta}
\end{equation}
Here, the value of \(\eta\) itself is supplied as a fitted, often tabulated, function of \(\lambda\) rather than derived from a constitutive principle. The boundary correction step then scales the trial deviatoric stress radially back onto the current surface \(\Delta\sigma(p,\lambda)\), preserving the trial stress ratios while the pressure is updated from the volumetric elastic response and \(\lambda\) is advanced for the next increment. 
The direction of this correction is set by the flow rule, which in KCC is deviatoric (and hence non-associated with respect to the pressure-dependent surface), with volumetric response supplied by a separate tabulated compaction relation.
Strain-rate sensitivity is incorporated through an additional empirical dynamic-increase-factor curve that further rescales the surfaces in Eq.~\eqref{eq:kcc_surfaces}.

The elastic-predictor, boundary-correction structure is not confined to single- or multi-surface plasticity models such as KCC. It also underlies constitutive frameworks that depart from a classical yield-surface formulation in stress-invariant space altogether. The microplane model M7 \citep{caner2013microplane1,caner2013microplane2} formulates the constitutive response on a set of predefined microplanes, planes of differing spatial orientation sampled over a unit hemisphere, rather than directly in terms of stress invariants. On each microplane, normal and shear strain components are related to corresponding stress components through fitted boundary functions that bound the admissible stress as a function of the volumetric, deviatoric, and shear strains on that plane. When the elastically predicted microplane stress vector exceeds its boundary, it is dropped back onto the boundary at constant strain, and the macroscopic stress is recovered from the ensemble of microplane stresses through the principle of virtual work. The admissible-stress region is thus defined plane-by-plane rather than through a single scalar surface, but the same elastic-trial-then-boundary-correction logic applies at the microplane level.

A related pattern appears in the Lattice Discrete Particle Model (LDPM) \citep{cusatis2011ldpm1,cusatis2011ldpm2}, a meso-scale formulation that represents concrete as an assembly of rigid particles connected through lattice facets rather than as a continuum. At each facet, an elastic law relates the normal and shear strain jumps between particles to facet tractions, and these tractions are in turn bounded by strength domains, for example fracturing, cohesive, and compaction/pore-collapse boundaries, defined as functions of the facet-level stress and strain state and fit to represent tensile, shear, and compressive failure mechanisms and their coupling. As with KCC and the microplane model, the elastic response is evaluated first, and the facet-level state is subsequently mapped back onto a boundary surface designed and calibrated in advance.

\subsection{The phenomenological cost of designing the boundary function}
\label{sec:phenomenological-cost}

Equations~\eqref{eq:kcc_surfaces} and~\eqref{eq:kcc_eta} exemplify what is implicit in most classical inelastic constitutive models: the elastic predictor in Eq.~\eqref{eq:elastic_trial} is essentially free, but the boundary function that performs the correction must be designed almost entirely from phenomenology. In the case of the KCC model, the three independent pressure-dependent surfaces, a tension/compression-sensitive damage-accumulation rule, a fitted interpolation function \(\eta(\lambda)\), and a separately calibrated rate-dependence curve together require a substantial battery of triaxial, uniaxial, and rate-dependent calibration tests, as well as engineering judgment about the functional forms in Eqs.~\eqref{eq:kcc_surfaces}--\eqref{eq:kcc_eta} themselves and all of the underlying rules. Once fitted, the resulting boundary is tied to the assumed rational and piecewise functional forms, and its predictions away from the calibration database depend on how well those assumed forms extrapolate rather than on a quantity learned directly from data or underlying mechanisms that are known to govern in the extrapolation region. Other constitutive frameworks replace the three KCC surfaces with a cap surface, a Drucker--Prager cone, a continuum damage variable, a set of microplane boundary functions, or lattice-facet strength domains, but the underlying task is the same: the admissible-stress boundary and its history dependence must be explicitly designed, using a chosen functional form and a calibration campaign suited to that form, before the elastic-predictor, boundary-correction update can be used.

This difficulty motivates the alternative approach pursued in this paper. Instead of committing in advance to a particular family of pressure-dependent strength surfaces and a fitted damage-interpolation function, the present model lets a latent state and a nonparametric Gaussian-process stress map absorb the analogous role, that of compressing irreversible loading history into quantities that modulate the current stress response directly from observed stress--strain sequences. The boundary-correction logic of Eqs.~\eqref{eq:kcc_surfaces} and~\eqref{eq:kcc_eta} is not imposed architecturally. Instead, the goal is to test whether a comparable pressure- and history-sensitive constitutive response can be learned through dependence on a small set of latent state variables, without prescribing the surfaces themselves.

In the terms of Section~\ref{sec:two-axes}, this places the present work in the data-driven paradigm. What distinguishes it within that paradigm is how much it deliberately retains from the phenomenological one: a set of internal variables that record loading history, an evolution law driven by a scalar measure of loading progress, and a stress response modulated by the current value of those variables. What it declines to retain is the closed-form specification of each of those ingredients. This approach is developed for broad classes of materials that are rate-independent and path-dependent, as described in Section~\ref{sec:formulation}. It is narrowed to quasi-brittle response only through the specific descriptor and irreversibility choices of Sections~\ref{sec:problem_qb} and~\ref{sec:formulation_qb}.

\section{Latent-state constitutive approach}
\label{sec:problem}

\subsection{Problem statement}

This section describes the learning problem, specifically defining the internal state variable and the associated evolution law that are inferred from observed stress--strain sequences as well as specifying the information that is assumed to be available for that inference.

The observable material response is represented using six-component strain and stress vectors,
\begin{equation}
\eps = [\varepsilon_{11},\varepsilon_{22},\varepsilon_{33},\varepsilon_{12},\varepsilon_{23},\varepsilon_{13}]^\top,
\qquad
\sig = [\sigma_{11},\sigma_{22},\sigma_{33},\sigma_{12},\sigma_{23},\sigma_{13}]^\top .
\end{equation}

Given a collection of stress--strain histories, the goal is to learn an incremental constitutive model
of the form
\begin{equation}
\sig_t = \mathcal{M}(\eps_t,\mathcal{H}_{t-1}),
\label{eqn:general_form}
\end{equation}
where \(\mathcal{H}_{t-1}\) denotes the unobserved material history accumulated before the current increment.

The only information assumed to be \textit{available} is a finite set of $N_s$ recorded stress--strain sequences \(\{(\eps^{(k)}_{1:T},\sig^{(k)}_{1:T})\}_{k=1}^{N_s}\), each consisting of $T$ load increments (the subscript $1{:}T$ denotes increments $1,\dots,T$), measured at the material-point scale, in which the applied strain path and the resulting stress response are both known at every step. 
Importantly, the following sources of information are \textit{not available}: the internal history \(\mathcal{H}\) is never measured directly; no functional form is prescribed for a strength surface, a flow rule, or a hardening law; and no structural-scale force--displacement data are used, so no such form can be back-calculated from structural response. The \textit{objective} is to learn an incremental model of the form in Eq.~\eqref{eqn:general_form} that is usable as a local constitutive update inside a finite element solver. That is, the model must be evaluated independently at each integration point using only information available at that point.

Classical constitutive models specify \(\mathcal{H}_{t-1}\) through chosen internal variables and evolution laws. In contrast, the present model represents this history through a learned latent state vector \(\z_t\in\mathbb{R}^{d_z}\),
\begin{equation}
\sig_t = \mathcal{M}(\eps_t,\z_t),
\qquad
\z_t = \mathcal{T}(\z_{t-1},\boldsymbol{\xi}_t).
\label{eq:gpssm_update}
\end{equation}
Here \(\boldsymbol{\xi}_t\) denotes a vector of frame-invariant scalar descriptors of the current increment, formed from the current strain and from the stress state at the end of the previous increment. Its role is to communicate how the material is currently being driven to the evolution law $\mathcal{T}$, in a form that does not depend on the orientation of the coordinate frame. The set of descriptors that enter \(\boldsymbol{\xi}_t\) is a material-specific modeling choice and is the one place where knowledge of the material class is used. The choice made for quasi-brittle materials is given in Section~\ref{sec:problem_qb}, and neither the argument of the present section nor the formulation of Section~\ref{sec:formulation} depends on that particular choice.

The latent state is not assumed to be a directly measurable physical quantity. It compresses the past loading path into a low-dimensional state needed for the current stress update. This is analogous to the role of internal variables in damage or plasticity theories, but with the representation learned from observable stress--strain sequences.

As with classical internal variables, the latent state \(\z_t\) is not uniquely identified by the training data: any invertible reparameterization of \(\z_t\), absorbed consistently into \(\mathcal{T}\) and \(\mathcal{M}\), reproduces the same stress predictions. The model is therefore identified only up to this functional equivalence class, which is sufficient for the constitutive-update role that \(\z_t\) plays, but also means that individual latent coordinates should not (without further evidence) be assigned a one-to-one physical interpretation such as a specific damage or plastic-strain mechanism.

Three questions follow immediately: 
\begin{enumerate}
    \item Why is a memoryless map from \(\eps_t\) to \(\sig_t\) insufficient?
    \item To what extent can a learned state vector legitimately replace the full loading path?
    \item Can such a latent learned state be interpreted mechanistically?
\end{enumerate}

The next two subsections address the first two questions in turn. However, the third question is a more difficult one (as alluded above). It is only addressed anecdotally here, and is left for further investigation in subsequent works that will aim to explore the physical basis of the latent vector and constraints that may be placed on it.  In Section~\ref{sec:problem_qb}, the specific choice of \(\boldsymbol{\xi}_t\) is then specified for the quasi-brittle materials studied here.

\subsection{The use of a latent state}

A state-space model is a commonly used tool in controls, system identification, and other fields to represents a dynamical system by an internal state that evolves over time and (together with the current input) determines both the next state and the current observation \citep{kalman1960,sarkka2013bayesian}. The central premise of this representation is that the observation alone is generally not sufficient to predict the future of the system: two systems can share an identical current observation and yet be in different states, and therefore evolve differently from that point onward.

Although historically it has not been posed in this way mathematically, the problem of constitutive modeling can be viewed through the state-space formalism.
Two material points can be observed at the same instant with an identical current strain \(\eps_t\), yet they may have accumulated different histories of damage, plastic strain, or micro-cracking at the micro-, meso-, or macroscale, over the loading paths that produced that strain. These internal quantities are not directly measurable in a standard stress--strain test, yet they are critical to understanding the evolution of the material with deformation.
A constitutive law therefore cannot be a memoryless map from strain to stress.
Of course this is well-understood in the materials modeling field. The distinction here is that classical constitutive theories supply this representation explicitly, through prescribed damage or plastic-strain variables. The state-space formulation pursued in this work instead treats it as an implicit state to be learned.

\subsection{Markov structure and probabilistic factorization}

The state-space formulation of Eq.~\eqref{eq:gpssm_update} is only meaningful if the constitutive update can be made Markovian in some state.  That is, there must exist a (possibly high-dimensional) history summary \(\mathcal{H}_t\) such that, conditioned on \(\mathcal{H}_t\), the future stress response is independent of the specific strain--stress path that produced \(\mathcal{H}_t\). Probabilistically, this is stated as follows.
Let \(\pi(\cdot)\) and
\(\pi(\cdot\mid\cdot)\) denote joint probability and conditional probability densities, and \(\z\sim\pi(\cdot)\) means that \(\z\) is distributed according to that density. The Markov property states that:
\begin{equation}
\pi\!\left(\sig_{t+1:T}\mid \eps_{1:T},\sig_{1:t},\mathcal{H}_t\right)=\pi\!\left(\sig_{t+1:T}\mid \eps_{t+1:T},\mathcal{H}_t\right).
\label{eq:markov}
\end{equation}
This is the same assumption that classical internal-variable constitutive theories make. Damage, plastic strain, and hardening variables are constructed so that the current stress increment can be computed from the current strain increment and the current values of these variables, without re-examining the entire loading history. It is also the assumption that underlies the Kalman filter and its nonlinear, Gaussian-process extensions \citep{kalman1960,wang2008gpdm,krishnan2017structured}. Replacing the full path with a state is only valid and useful if that state is Markovian. 

Here, the learned latent state \(\z_t\) is constructed to satisfy Eq.~\eqref{eq:markov} by design, since Eq.~\eqref{eq:gpssm_update} defines \(\z_t\) recursively from \(\z_{t-1}\) and the current-step descriptors \(\boldsymbol{\xi}_t\), and never from the strain--stress history directly. It follows that \(\z_t\) is Markovian by construction rather than by empirical verification, and its adequacy as a finite-dimensional summary of \(\mathcal{H}_t\) is tested in the verification cases of Section~\ref{sec:results}. This Markov structure also makes the finite element deployment of Section~\ref{sec:data} practical. Because \(\z_t\) (together with the previous-step stress from which \(\boldsymbol{\xi}_t\) is formed) is sufficient to continue the update, it can be stored as a small, fixed-size solver state variable at each integration point, rather than requiring the full strain history to be retained and replayed at every increment.

Given this Markov state, the model factorizes 
into a transition step, which governs the evolution of the latent step,
and an observation step, which specifies the conditional distribution of $\sig_t$ given the current state:
\begin{equation}
\z_t \sim \pi(\z_t \mid \z_{t-1}, \boldsymbol{\xi}_t),
\qquad
\sig_t \sim \pi(\sig_t \mid \z_t,\eps_t).
\end{equation}
The nominal transition used for inference and finite element deployment is deterministic given the descriptors, \(\z_t=\mathcal{T}(\z_{t-1},\boldsymbol{\xi}_t)\) as in Eq.~\eqref{eq:gpssm_update}. During training, increment-scaled Gaussian perturbations are added to this transition and marginalized by Monte Carlo rollout, as detailed in Section~\ref{sec:formulation}. This regularizes the learned state trajectory but is not propagated in the deployed constitutive update. Only the Gaussian-process stress map requires an inducing-point posterior approximation, in which a small set of representative points summarizes the training data so that the cost of evaluating the stress map does not grow with the number of observations.
The model therefore does not, in its present form, quantify epistemic uncertainty in the learned history representation itself, only uncertainty in the stress prediction conditioned on that history. The transition parameters are point estimates, and the state noise introduced during training is a fixed learned quantity rather than a posterior.

\subsection{Specialization to quasi-brittle materials: invariant descriptors}
\label{sec:problem_qb}

The setting and the state-space argument above are generic. Nothing in Eq.~\eqref{eq:gpssm_update} or Eq.~\eqref{eq:markov} is specific to a material class. The model becomes material specific when we choose the scalar descriptors that make up \(\boldsymbol{\xi}_t\) and drive the latent transition. The choices made below specialize the general setting to isotropic quasi-brittle materials under triaxial loading, where pressure sensitivity and deviatoric stress intensity are the dominant drivers of the inelastic response.

A scalar equivalent strain is used to measure loading progression, defined as
\begin{equation}
\eeq = \sqrt{\frac{2}{3}\left[(\varepsilon_{11}-\varepsilon_{22})^2+(\varepsilon_{22}-\varepsilon_{33})^2+(\varepsilon_{33}-\varepsilon_{11})^2+6(\varepsilon_{12}^2+\varepsilon_{23}^2+\varepsilon_{13}^2)\right]} .
\label{eq:eeq}
\end{equation}
To encode pressure sensitivity and deviatoric stress intensity, the previous-step stress state is summarized using the first stress invariant and the second invariant of the deviatoric stress, given by
\begin{equation}
I_1=\tr(\sig)=\sigma_{11}+\sigma_{22}+\sigma_{33},
\end{equation}
\begin{equation}
J_2=\frac{1}{2}\sdev:\sdev,
\qquad
\sdev=\sig-\frac{1}{3}I_1\I .
\end{equation}
In the six-component convention used here,
\begin{equation}
J_2=\frac{1}{6}\left[(\sigma_{11}-\sigma_{22})^2+(\sigma_{22}-\sigma_{33})^2+(\sigma_{33}-\sigma_{11})^2\right]
+\sigma_{12}^2+\sigma_{23}^2+\sigma_{13}^2 .
\end{equation}
The descriptor vector of Eq.~\eqref{eq:gpssm_update} is then taken to be
\begin{equation}
\boldsymbol{\xi}_t=\left(\eeq_{t},\,I_{1,t-1},\,J_{2,t-1}\right),
\label{eq:xi_qb}
\end{equation}
so that the latent transition sees the current loading progression together with the pressure and deviatoric intensity of the stress state it starts from. These descriptors do not replace the full tensorial stress--strain relation. Instead, they provide compact mechanics-informed inputs for latent-state evolution.

The choice of \(I_1\) and \(J_2\) is not arbitrary. Rather, it mirrors the pressure--deviatoric split that underlies the boundary functions of Section~\ref{sec:background}. The mean stress \(p=I_1/3\) controls confinement-dependent strength and stiffness in exactly the way the yield, maximum, and residual surfaces \(\Delta\sigma_y(p)\), \(\Delta\sigma_m(p)\), \(\Delta\sigma_r(p)\) of Eq.~\eqref{eq:kcc_surfaces} are parameterized by pressure, while \(\sqrt{3J_2}\), a von-Mises equivalent stress, measures the distortional stress intensity that drives shear-dominated microcracking and is the quantity actually compared against those surfaces. Feeding \(I_1\) and \(J_2\) into the latent transition therefore gives the learned history variable access to the same two stress descriptors that a classical pressure-dependent plasticity model would use to evaluate its yield function, without committing to the specific rational functional forms of e.g. Eqs.~\eqref{eq:kcc_surfaces}--\eqref{eq:kcc_eta}. 

The equivalent strain \(\eeq\) of Eq.~\eqref{eq:eeq} is an explicit state variable. It is an accumulating scalar measures of loading progression used to drive history evolution, that is used in lieu of the full strain tensor. Together with the state vector $\z_t$, it plays a role analogous to the the scalar damage-like history variable \(\lambda\) of Eq.~\eqref{eq:kcc_eta}. 
Unlike \(\lambda\), which evolves through a pressure dependent function $h(p)$ acting on the evolution of effective plastic strain as
\begin{equation}
    \dot{\lambda} = h(p)\dot{\bar{\varepsilon}}^p
    \label{eqn:KCC_damage_evolution}
\end{equation}
where $\dot{\bar{\varepsilon}}^p=\sqrt{2/3\dot{\boldsymbol{\varepsilon}}^p:\dot{\boldsymbol{\varepsilon}}^p}$ and $h(p)$ requires calibration, 
\(\eeq\) is defined directly from the strain tensor and it deliberately excludes the volumetric strain component. 
For quasi-brittle materials under triaxial compression, volumetric and deviatoric strain are governed by different mechanisms: volumetric strain reflects competing pore collapse and microcrack dilatancy, whereas deviatoric strain grows steadily through frictional sliding and crack growth.
The pressure channel through which volumetric information enters the model is instead carried by the stress invariant \(I_1\), consistent with the classical treatment in stress-space rather than a strain-space quantity.

During supervised training, previous-step invariants may be computed from measured stresses to stabilize the latent rollout and expose the transition to physically meaningful pressure and deviatoric descriptors. During finite element deployment, measured stresses are unavailable and the same invariants are recomputed from the model-predicted stress at the previous increment. The deployed update is therefore fully local and self-contained.

Violating frame invariance is a recognized failure mode of neural-network constitutive models. Without architectural or data-level safeguards, such models can learn spurious dependence on the arbitrary orientation of the material coordinate frame, so that the predicted stress changes when the same deformation is merely expressed in a rotated frame \citep{heider2020so3,vlassis2020geometric}. The present model addresses this concern in two complementary ways. First, the latent transition depends only on the rotation-invariant descriptors \(\eeq\), \(I_1\), and \(J_2\). Second, the Gaussian-process stress map is trained under SO(3) rotational data augmentation (Section~\ref{sec:data}).

\section{Gaussian-process state-space constitutive formulation}
\label{sec:formulation}

As one instance of this structure, the implementation developed here evolves the latent variable with a deterministic neural map and predicts stress with a Gaussian process, so as to balance speed with the need for uncertainty quantification. The transition is evaluated at every integration point at every increment and is kept cheap, while the stress map retains a posterior variance that provides an epistemic measure for where the training data do and do not constrain the prediction.

\subsection{Latent state evolution}

The constitutive model is written as a discrete state-space model,
\begin{equation}
\sig_t=g(\z_t,\eps_t)+\mathbf{v}_t,
\qquad
\z_t=\z_{t-1}+\Delta\z_t,
\label{eq:state_space_noise}
\end{equation}
where \(g\) is a probabilistic stress map and \(\mathbf{v}_t\) is a discrepancy (or noise) term.
The latent-state increment, $\Delta\z_t$, is parameterized by a neural transition function of the latent state together with two ingredients that describe how the material is being driven. The first is a scalar loading-progress parameter \(\kappa\), which measures the increment of deformation applied in a given step.
The second is the frame-invariant descriptor vector \(\boldsymbol{\xi}_t\) introduced in Section~\ref{sec:problem}, which summarizes the conditions under which the current increment is applied. The transition is then modeled by
\begin{equation}
\Delta\z_t=
f_\theta\left[\z_{t-1},\boldsymbol{\xi}_t\right]
\left|\Delta\tilde{\kappa}_t\right|,
\label{eq:latent_transition}
\end{equation}
where tildes denote normalized quantities and
\begin{equation}
\Delta\tilde{\kappa}_{t}=\frac{\kappa_t-\kappa_{t-1}}{s_{\kappa}} ,
\label{eq:kappa_increment}
\end{equation}
with \(s_{\kappa}\) a fixed scale constant. Note the similarity between the discrete form in Eq.~\eqref{eq:latent_transition} and the continuous damage evolution law in Eq.~\eqref{eqn:KCC_damage_evolution}. However, while Eq.~\eqref{eqn:KCC_damage_evolution} is strongly prescriptive, the proposed state transition function is general and highly flexible. Indeed, the progress parameter \(\kappa\), the descriptor vector \(\boldsymbol{\xi}\), the dimension of the state vector $\z_t$, and any constraint placed on the output of \(f_\theta\) are not specified here. They are material-dependent and can be customized to account for the specific mechanisms, constraints, and data available for a given material. For quasi-brittle materials, these are assigned in Section~\ref{sec:formulation_qb}.

The rate-times-increment form in Eq.~\eqref{eq:latent_transition} encodes the central inductive bias of the formulation. That is, the neural network learns a rate of latent variable evolution per unit of deformation, not an increment. Because the update is scaled by \(|\Delta\tilde{\kappa}_t|\), the latent state advances only when deformation progresses, and when the increment in \(\kappa\) is nearly zero the same factor suppresses artificial latent drift. Nothing in Eq.~\eqref{eq:latent_transition} restricts the sign of that rate, so a history variable that can decrease is admissible. Instead, material-specific knowledge about which evolutions are admissible is injected by constraining the output of \(f_\theta\), and Section~\ref{sec:formulation_qb} does exactly that for the material class studied here.

Eq.~\eqref{eq:latent_transition} is a general form, but it is not a universal one. First, it assumes \emph{rate independence}. Evolution is parameterized by \(\kappa\) rather than by time, so two loadings that trace the same path at different speeds produce the same response. A rate-dependent material, whether through viscoplasticity or through strain-rate-sensitive strength, would require replacing \(\kappa\) by a time-parameterized deformation measure, at the cost of the sampling-rate invariance described next. Second, it assumes that \(\boldsymbol{\xi}_t\) is \emph{frame invariant}. If any component of \(\boldsymbol{\xi}_t\) changes under a rigid rotation of the coordinate frame, so would the predicted latent evolution, and the constitutive update would inherit a dependence on an arbitrary choice of material axes. Together these two assumptions restrict the present formulation to rate-independent, path-dependent materials. Within that class, the remaining physics enters through the choice of \(\kappa\), \(\boldsymbol{\xi}\), the dimension of $\z_t$, and the constraints placed on \(f_\theta\).  It is at this level, not in Eq.~\eqref{eq:latent_transition} itself, that our approach commits to a particular material class.

The rate-times-increment factorization also decouples the learned state-evolution tendency from the sampling interval of the training data. The network \(f_\theta\) predicts a rate per unit of loading progress, while \(|\Delta\tilde{\kappa}_t|\) rescales that rate to the increment actually observed, so the same trained transition remains valid whether a loading path is sampled finely or coarsely. This is particularly useful for finite element deployment, where the solver increment is set by explicit stability and mesh constraints rather than by the experimental sampling interval, yet requires no retraining of the transition to accommodate.

Writing the update in the continuum limit of small increments gives an ordinary differential equation in deformation progress rather than in real time,
\begin{equation}
\frac{d\z}{d\kappa} = f_\theta\left[\z,\boldsymbol{\xi}\right]\operatorname{sign}(\Delta\kappa),
\label{eq:latent_ode}
\end{equation}
which makes explicit that the latent evolution is parameterized by \(\kappa\) rather than by wall-clock or explicit-solver time.

\subsection{Specialization to quasi-brittle materials}
\label{sec:formulation_qb}

For the isotropic and rate-independent quasi-brittle materials studied here, the material-dependent ingredients of Eq.~\eqref{eq:latent_transition} are assigned as
\begin{equation}
\kappa=\eeq,
\qquad
\boldsymbol{\xi}_t=\left(\eeqt_t,\tilde{I}_{1,t-1},\tilde{J}_{2,t-1}\right),
\qquad
f_\theta\ge\mathbf{0},
\label{eq:qb_assignment}
\end{equation}
with \(\eeq\) the equivalent strain of Eq.~\eqref{eq:eeq} and \(I_1\), \(J_2\) are the stress invariants of Section~\ref{sec:problem_qb}, each entering the network in its normalized form. 
Equivalent strain is a natural loading-progress parameter for this material class because the quasi-static triaxial data used for training are rate independent, so deformation rather than time is the meaningful measure of how far a loading path has advanced.

The nonnegativity of \(f_\theta\) is imposed through a softplus output nonlinearity, and it represents the material-specific inductive bias of the present application. It makes \(\Delta\z_t\ge\mathbf{0}\) at every increment and, in the continuum limit of Eq.~\eqref{eq:latent_ode}, it makes \(d\z/d\kappa\ge\mathbf{0}\) as loading progresses, so the latent state accumulates along a loading path and never reverses. This is appropriate here because the mechanisms the latent state stands in for, namely damage, compaction, and microcrack growth, only accumulate. This assumption would not be appropriate for a material whose history variable must be able to decrease, such as kinematic hardening and the associated Bauschinger effect, or porosity. Eq.~\eqref{eq:latent_transition} admits such materials; only the sign constraint imposed here excludes them.

In the present implementation, the transition network is a multilayer perceptron with input dimension five and output dimension \(d_z=2\). It uses three hidden layers of width 64 with hyperbolic-tangent activations, followed by the softplus output nonlinearity. The scalar descriptors \(\eeq\), \(I_1\), and \(J_2\) are invariant under rigid coordinate rotations, so the state-transition rate is not tied to a particular material-axis orientation.

Mechanically, Eq.~\eqref{eq:latent_transition} plays the role that a flow rule and hardening law play in classical rate-independent plasticity. The plastic multiplier is replaced by the equivalent-strain increment \(|\Delta\eeqt_t|\), and the flow direction and magnitude are replaced by the learned, invariant-dependent nonnegative rate \(f_\theta[\cdot]\). The hyperbolic-tangent hidden-layer activations keep \(f_\theta\) smooth and bounded, which in turn keeps the right-hand side of Eq.~\eqref{eq:latent_ode} Lipschitz in \(\z\). This is a modeling choice rather than an enforced constraint, but it favors a well-behaved, non-stiff latent trajectory when Eq.~\eqref{eq:latent_transition} is integrated explicitly at the small time increments used by the explicit finite element solver.

\subsection{Sparse Gaussian-process stress map}

The stress predictor uses the combined normalized input
\begin{equation}
\x_t=[\z_t,\tilde{\eps}_t]\in\mathbb{R}^{d_z+6} .
\end{equation}
Feeding a jointly trained representation into a Gaussian-process kernel in this way follows the deep-kernel-learning precedent, in which a neural network produces the feature space over which a GP operates \citep{wilson2016deep}. The difference is that \(\z_t\) here is not a fixed embedding of a single input but a recurrent latent state that retains information about preceding loading increments.

Each stress component is modeled by a Gaussian process with a radial basis function (RBF) kernel and automatic relevance determination (ARD),
\begin{equation}
k_a(\x,\x')=\alpha_a^2\exp\left[-\frac{1}{2}\sum_{r=1}^{d_z+6}\frac{(x_r-x_r')^2}{\ell_{a,r}^2}\right],
\qquad a=1,\ldots,6 .
\end{equation}
The ARD length scales \(\ell_{a,r}\) give each output its own sensitivity to every latent-state and strain coordinate. Small \(\ell_{a,r}\) implies rapid variation of stress component \(a\) along input direction \(r\), while large \(\ell_{a,r}\) implies weak dependence over the range represented by the data.
The six latent GPs have separate means, kernels, and variational distributions and are conditionally independent given \(\x_t\).
They remain coupled through their common dependence on the latent history \(\z_t\) and current strain \(\eps_t\), but the output layer contains no explicit cross-output covariance or coregionalization kernel \citep{alvarez2012kernels}.
This choice keeps the six posterior-mean evaluations independent at deployment, at the cost of not representing residual cross-component covariance beyond the dependence already expressed by \(\x_t\).

Exact GP regression with a Gaussian likelihood has a closed-form posterior, but factorizing the covariance matrix over \(N\) training points requires \(\mathcal{O}(N^3)\) operations and \(\mathcal{O}(N^2)\) storage.
We instead use a sparse variational GP (SVGP) with \(M=150\) inducing inputs per stress component, where \(M\ll N\) \citep{titsias2009variational,hensman2013big}.

The six outputs share a single set of inducing locations
\(\mathbf{Z}\in\mathbb{R}^{M\times(d_z+6)}\), optimized jointly with the remaining variational
parameters, while each output retains its own kernel hyperparameters, mean, and variational
distribution. Sharing \(\mathbf{Z}\) is deliberate. The constitutive update
then evaluates one \(M\)-point kernel distance table per material point and reuses it across all
six stress components, and the exported parameter file contains one \(M\times(d_z+6)\) block of
inducing coordinates rather than six.

For component \(a\), let $\mathbf{u}_a=f_a(\mathbf{Z})$
denote the stress component evaluated by the GP at the shared inducing points \(\mathbf{Z}\), $\pi_0(\mathbf{u}_a)\sim\mathcal{N}(\mathbf{0},\mathbf{K}_{ZZ,a})$ denote the prior distribution on $\mathbf{u}_a$, and $q(\mathbf{u}_a)\sim\mathcal{N}(\mathbf{m}_a,\mathbf{S}_a)$ denote the variational posterior distribution of $\mathbf{u}_a$. Here,  \(\mathbf{K}_{ZZ,a}=k_a(\mathbf{Z},\mathbf{Z})\) is formed with the kernel hyperparameters of
output \(a\), and the mean $\mathbf{m}_a$ and Cholesky factor of \(\mathbf{S}_a\) are learned.
The variational marginal at a training input, indexed by $i$, is obtained by retaining the GP prior conditional and integrating over the inducing-variable posterior,
\begin{equation}
q(f_{ia})
=\int \pi_0(f_{ia}\mid\mathbf{u}_a)\,
q(\mathbf{u}_a)\,d\mathbf{u}_a, \qquad i=1,\dots, N .
\end{equation}
Thus, the inducing variables provide a low-dimensional parameterization of the approximate function posterior rather than replacing the GP by values defined only at \(\mathbf{Z}\) \citep{matthews2016sparse}.

For normalized stress observation \(y_{ia}\), the six-output evidence lower bound (ELBO) is given by
\begin{equation}
\mathcal{L}_{\mathrm{ELBO}}
=
\sum_{a=1}^{6}
\left\{
\sum_{i=1}^{N}
\mathbb{E}_{q(f_{ia})}
\left[\log \pi(y_{ia}\mid f_{ia})\right]
-
\mathrm{KL}
\left[q(\mathbf{u}_a)\|\pi_0(\mathbf{u}_a)\right]
\right\}.
\label{eq:svgp_elbo}
\end{equation}
where the log likelihood, $\log \pi(y_{ia}\mid f_{ia})$, measures how well the predictive function distribution explains the observed stresses, and the Kullback-Leibler divergence (KL) penalizes unsupported departure of the inducing-variable posterior from the GP prior.
Maximizing Eq.~\eqref{eq:svgp_elbo} therefore balances probabilistic stress fitting against posterior complexity rather than minimizing a deterministic mean-squared error alone \citep{blei2017variational}.
For the Gaussian observation model
\begin{equation}
\pi(y_{ia}\mid f_{ia})
\sim\mathcal{N}(y_{ia};f_{ia},\tau_a^2),
\end{equation}
the expected log likelihood can be evaluated analytically as:
\begin{equation}
\mathbb{E}_{q(f_{ia})}
\left[\log \pi(y_{ia}\mid f_{ia})\right]
=-\frac{1}{2}
\left[
\frac{(y_{ia}-\mu_{ia})^2+v_{ia}}{\tau_a^2}
+\log(2\pi\tau_a^2)
\right],
\label{eqn:ELBO_loglike}
\end{equation}
where \(\mu_{ia}\) and \(v_{ia}\) are the variational mean and variance of \(f_{ia}\), and \(\tau_a^2\) is the learned observation-noise variance in normalized stress units.
Consequently, the likelihood term accounts for both mean error and function uncertainty.

Training uses the normalized form of the GPyTorch~\citep{gardner2018gpytorch} variational objective, evaluated over the
full set of training points at every optimization step.
Writing \(\mathcal{B}\) for the set of evaluated points and \(B=|\mathcal{B}|\), the loss function can be written as
\begin{equation}
\widehat{\mathcal{L}}_{\mathcal{B}}
=
\frac{1}{B}
\sum_{i\in\mathcal{B}}\sum_{a=1}^{6}
\mathbb{E}_{q(f_{ia})}
\left[\log \pi(y_{ia}\mid f_{ia})\right]
-
\frac{1}{N_{\mathrm{phys}}}
\sum_{a=1}^{6}
\mathrm{KL}
\left[q(\mathbf{u}_a)\|\pi(\mathbf{u}_a)\right].
\label{eq:minibatch_elbo}
\end{equation}
where \(N_{\mathrm{phys}}\)
is the number of original physical training points rather than the number after rotational augmentation.
Rotated copies are not independent experiments. Counting them as additional physical observations in the KL term would artificially weaken the prior regularization.

In the present study, a single optimization step evaluates all \(B\) augmented training points,
so no stochastic subsampling of the data is used. The normalization above is written in the
general form because it is unchanged if \(\mathcal{B}\) is instead a random subset (i.e.\ mini-batching), thus supporting a sparse formulation for larger dataset sizes.

The latent variable transition model and the GP stress model are trained concurrently. A Monte Carlo average is used for the stochastic latent-state rollout such that,
during training, the transition increment is perturbed as
\begin{equation}
\z_t^{(s)}
=\z_{t-1}^{(s)}
+\Delta\z_t^{\mathrm{mean}}
+\sigma_z\left|\Delta\eeqt_t\right|\mathbf{w}_t^{(s)},
\qquad
\mathbf{w}_t^{(s)}\sim\mathcal{N}(\mathbf{0},\mathbf{I}),
\label{eq:stochastic_rollout}
\end{equation}
and the minimized loss is
\begin{equation}
\mathcal{L}_{\mathrm{train}}
=-\frac{1}{S}\sum_{s=1}^{S}
\widehat{\mathcal{L}}_{\mathcal{B}}
\left(\z^{(s)}\right),
\qquad S=3 .
\label{eq:total_training_loss}
\end{equation}
Because the sampled transition is reparameterized and the GP input depends on \(\z_t^{(s)}\), gradients of Eq.~\eqref{eq:total_training_loss} propagate through the entire rollout \citep{kingma2014autoencoding}.
A single optimization step therefore updates the transition-network parameters and its learned noise scale, the inducing locations and variational moments, the kernel and mean parameters, and the likelihood noise.
For a full pass through the data, the leading SVGP cost is \(\mathcal{O}(NM^2+M^3)\) instead of
the \(\mathcal{O}(N^3)\) cost of exact GP training; restricting \(\mathcal{B}\) to a subset of
\(B\) points would reduce this to \(\mathcal{O}(BM^2+M^3)\) per step.

For finite element implementation, the trained GP is exported in posterior-mean form,
\begin{equation}
\mu_a(\x)=c_a+\mathbf{k}_a(\x,\mathbf{Z})^\top\boldsymbol{\beta}_a,
\label{eq:gp_mean}
\end{equation}
where \(c_a\) is the learned mean term and \(\boldsymbol{\beta}_a\) is precomputed from the trained variational parameters.
Training uses the whitened representation \(\mathbf{u}_a=\mathbf{L}_a\mathbf{u}_{w,a}\), with \(\mathbf{u}_{w,a}\sim\mathcal{N}(\mathbf{0},\mathbf{I})\) under the prior, to improve conditioning and simplify evaluation of the KL term \citep{matthews2016sparse}.
For the whitened SVGP,
\begin{equation}
\boldsymbol{\beta}_a=\mathbf{L}_a^{-\top}\mathbf{m}_{w,a},
\qquad
\mathbf{L}_a\mathbf{L}_a^\top=\mathbf{K}_{ZZ,a},
\end{equation}
where \(\mathbf{m}_{w,a}\) is the learned whitened variational mean.
The deployed predictor therefore requires only \(M\) kernel evaluations and one dot product per stress component. Neither latent-state sampling nor posterior covariance propagation is used in the VUMAT.

Equation~\eqref{eq:gp_mean} can be read mechanically as a learned, nonparametric generalization of a material-response lookup table. The predicted stress at a given latent state and strain is a kernel-weighted combination of representative points in the joint history--strain space, with weights that decay according to the learned ARD metric.
This plays a role analogous to the piecewise, pressure-indexed strength surfaces of Eq.~\eqref{eq:kcc_surfaces}, but the interpolation structure and its effective resolution are learned from data rather than prescribed through a fixed rational functional form.

The six posterior means of Eq.~\eqref{eq:gp_mean},  evaluated at the current input \(\x_t\), form the normalized stress vector
\(\tilde{\sig}_t=[\mu_1(\x_t),\ldots,\mu_6(\x_t)]^\top\), which is transformed back to physical stress units by
\begin{equation}
\sig_t=\tilde{\sig}_t s_\sigma+\boldsymbol{\mu}_\sigma,
\label{eq:denormalize_stress}
\end{equation}
where \(\boldsymbol{\mu}_\sigma\) and \(s_\sigma\) are the component-wise stress mean and shared scalar stress scale, respectively, used for normalizing all stress components, following the conventions of Section~\ref{sec:norm}.

\subsection{Role of uncertainty}
\label{sec:uncertainty-decomposition}
The GP observation model defines both a posterior mean and a posterior variance. This is important because limited triaxial data (which is what we use in the training) cannot cover all loading paths that may arise in a finite element analysis. The probabilistic structure can therefore be used offline to identify data-sparse regions, compare model variants, facilitate uncertainty propagation, and guide active learning. This is a direct statistical countermeasure to the phenomenological limitation identified in Section~\ref{sec:background}. For example, the rational functions of Eq.~\eqref{eq:kcc_surfaces} extrapolate according to a fixed, assumed functional form and cannot signal when a prediction has left the calibration regime, whereas the GP posterior variance grows, in principle, in regions of the joint latent-state/strain space that are far from the inducing points and training data. This provides an inbuilt, quantitative indicator of interpolation/extrapolation risk rather than relying on engineering judgment about functional-form validity.

The GP can also account for noise or inherent stochastic variability in the material behavior and the associated training data. In fact, the probabilistic structure retained by the sparse variational formulation allows this predicted stress uncertainty to be separated into contributions of distinct origin. Because the latent state enters the stress map as a random variable, the predictive variance follows from the law of total variance,
\begin{equation}
\label{eq:total-variance}
\underbrace{\operatorname{Var}\!\left[\boldsymbol{\sigma}_t \mid
  \boldsymbol{\varepsilon}_{1:t}\right]}_{\text{total}}
=
\underbrace{\mathbb{E}_{\mathbf{z}_t}\!\left[
  \operatorname{Var}_{\mathrm{GP}}\!\left(\boldsymbol{\sigma}_t \mid
  \mathbf{z}_t, \boldsymbol{\varepsilon}_t\right)\right]}_{\displaystyle
  \sigma^2_{\mathrm{GP}}}
+
\underbrace{\operatorname{Var}_{\mathbf{z}_t}\!\left[
  \mathbb{E}_{\mathrm{GP}}\!\left(\boldsymbol{\sigma}_t \mid
  \mathbf{z}_t, \boldsymbol{\varepsilon}_t\right)\right]}_{\displaystyle
  \sigma^2_{\mathbf{z}}}
+
\underbrace{\sigma^2_{n}}_{\text{likelihood}},
\end{equation}
evaluated component-wise on the six stress components, where the first term denotes the epistemic variance of the stress prediction, the second term relates to the aleatory stochasticity in the latent state evolution, and the final term is a stochastic noise. In the implementation, the outer expectation is taken over \(S_{\mathrm{eval}}=100\) Monte Carlo rollouts of the latent trajectory, each propagated from $\mathbf{z}_0 = \mathbf{0}$ through the stochastic transition of Eq.~\eqref{eq:stochastic_rollout}, with the sparse GP posterior evaluated along every realisation. This is deliberately larger than the \(S=3\) samples used during training when evaluating the loss of Eq.~\eqref{eq:total_training_loss}. During training, the average only has to supply an
unbiased gradient, whereas the variance decomposition reported during model evaluation requires a converged estimate
of each individual term.

The three terms have distinct physical interpretations. The epistemic term, $\sigma^2_{\mathrm{GP}}$, is the posterior variance of the sparse GP at the query point $(\mathbf{z}_t, \boldsymbol{\varepsilon}_t)$, averaged over the sampled latent trajectories. It measures how far the current state--strain pair lies from the inducing-point support built during training, and therefore reports the degree to which the model is extrapolating. It is the only term that is reducible by performing additional experiments. Collecting a new loading history that covers a previously unseen region of state--strain space reduces this term. The remaining two terms, by contrast, cannot be reduced by adding data. The latent-induced term, $\sigma^2_{\mathbf{z}}$, arises because the transition network carries a learnable state noise, the increment-scaled perturbation
$\sigma_z|\Delta\eeqt_t|\mathbf{w}_t^{(s)}$ of Eq.~\eqref{eq:stochastic_rollout}, so that different rollouts reach slightly different states at the same strain. It is the resulting spread of the GP posterior \emph{mean} due to randomness in the latent state, and quantifies how strongly the stress prediction is influenced by uncertainty in the inferred history variable. It will grow, for example, if the learned state provides a highly variable summary of the latent history, which might occur when the stress-strain response of the material is highly variable under nominally identical conditions. The aleatoric term $\sigma^2_{n}$ is the learned homoscedastic noise of the multitask Gaussian likelihood, and is constant along a sequence and across sequences. All three are computed in the normalized stress units used for training and recovered in physical units through the scaling in Eq.~\eqref{eq:denormalize_stress}.

Since the aleatoric terms are essentially the same for every sequence, they do not change how sequences compare. That is, ranking candidate experiments by the total predictive standard deviation or by the epistemic term alone gives the same order.
Reporting the split, rather than the total alone makes the distinction available to the user for actions such as active learning. A fixed-form constitutive model calibrated by least squares supplies a single parameter set and no comparable internal measure of where its calibration is unsupported. In the present VUMAT implementation, only the posterior mean in Eq.~\eqref{eq:gp_mean} is evaluated for efficiency and robustness. Thus, the finite element results reported below should be interpreted as posterior-mean deployment results, not as a full uncertainty-propagating finite element analysis -- which the proposed framework enables but are left for future work.

\subsection{Framework overview}

\begin{figure}[!ht]
    \centering
    \includegraphics[width=0.8\textwidth]{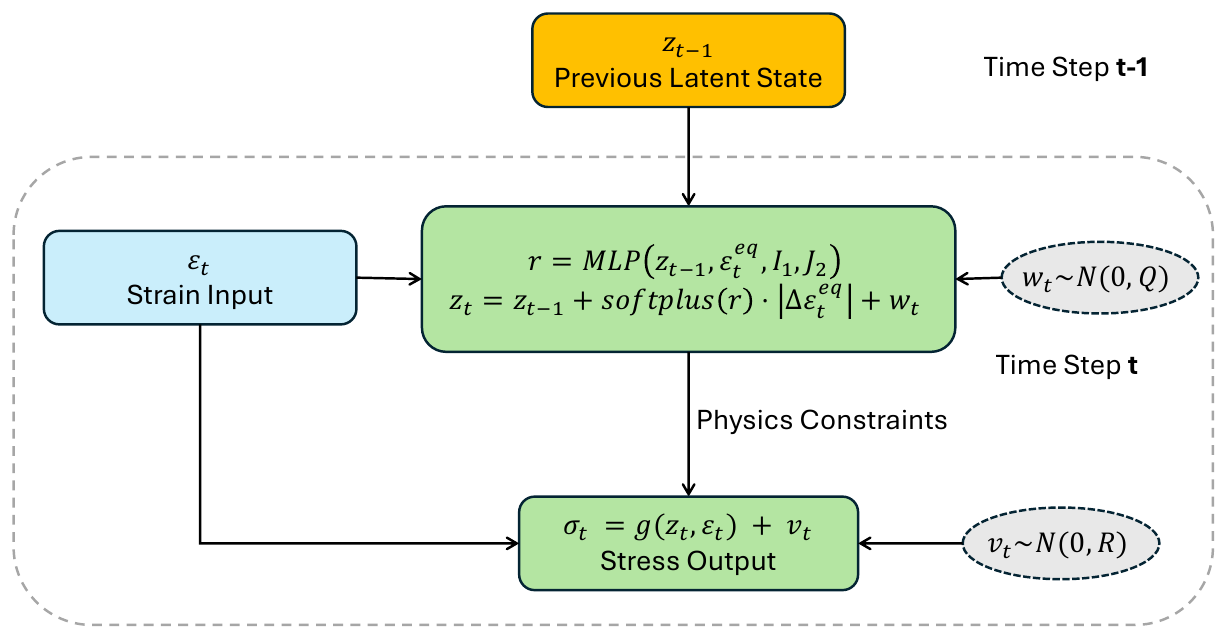}
    \caption{Schematic of the latent-state transition and stress emission in the GP-SSM constitutive update: the latent state \(\z_{t-1}\) is advanced through the mechanics-informed transition of Eq.~\eqref{eq:latent_transition} to obtain \(\z_t\), which together with the current strain \(\eps_t\) is mapped by the Gaussian-process stress map \(g\) to the stress \(\sig_t\). 
    }
    \label{fig:gpssm_schematic}
\end{figure}

Figure~\ref{fig:gpssm_schematic} summarizes the incremental constitutive update. At each time step \(t\), the current strain \(\eps_t\) is used to calculate the equivalent strain and its increment. Together with the previous latent state \(\z_{t-1}\) and the previous-step stress invariants \(I_{1,t-1}\) and \(J_{2,t-1}\), these inputs determine the latent evolution rate through a multilayer perceptron. A softplus transformation makes the deterministic rate nonnegative, and multiplication by the magnitude of the normalized equivalent-strain increment gives the latent-state increment in Eq.~\eqref{eq:latent_transition}. The updated state \(\z_t\) is then combined with the current strain and passed to the sparse GP stress map to predict the six stress components. The terms \(\mathbf{w}_t\) and \(\mathbf{v}_t\) in the schematic represent latent-transition and observation noise, respectively; the implemented transition noise is scaled by the equivalent-strain increment as in Eq.~\eqref{eq:stochastic_rollout}. During closed-loop prediction, the predicted stress supplies the invariants for the next step, while \(\z_t\) carries the accumulated loading history forward. The transition network and GP stress map are trained jointly, whereas finite element deployment evaluates the deterministic latent update and GP posterior mean.

\section{Data Processing and Training}
\label{sec:data}

\subsection{Stress--strain histories and normalization}
\label{sec:norm}

Training begins by providing a set of $N_{\text{hist}}$ triaxial stress-strain histories separated into an $N_{\text{train}}/N_{\text{val}}/N_{\text{test}}$ training/validation/test sequences, with the exact counts used in this study provided in \ref{sec:implementation}. Each sequence is preprocessed by removing the initial transient segment, inserting a zero reference state, truncating by equivalent strain, and normalizing strain, stress, equivalent strain, \(I_1\), and \(J_2\). The zero reference state gives the recurrent model a common undeformed initial condition, while equivalent-strain truncation prevents the training objective from being dominated by extreme extrapolation regions outside the intended calibration domain.

The normalization convention for a quantity $\xi$ is given by
\begin{equation}
\tilde{\xi}=\frac{\xi-\mu_\xi}{s_\xi} .
\end{equation}
A shared scalar standard deviation is used for all six strain components, and another shared scalar standard deviation is used for all six stress components, rather than normalizing each component independently. This choice is important for rotational augmentation. Because the original triaxial data contain nearly zero shear components, component-wise normalization would artificially amplify rotated shear components. Shared tensor-component scaling instead preserves the relative structure of the stress and strain tensors under rotation. The scalar descriptors \(\eeq\), \(I_1\), and \(J_2\) are normalized independently, since they are already rotation-invariant scalars and do not share this constraint. The specific normalization statistics fitted for the exported model used in the present verification are reported in \ref{sec:statistics}.

\subsection{Temporal downsampling}
\label{sec:downsample}

Because the model is trained directly on experimentally measurable quantities (the stress and strain histories themselves, rather than on a quantity defined at a fixed clock rate), it must also accommodate the fact that different triaxial tests, and different finite element solves, are not recorded or advanced at a common temporal resolution. One experiment may report tens of points along a loading path, another many thousands, and neither sampling rate is dictated by the material response itself. A constitutive model that learned a rate with respect to wall-clock or step time would inherit this arbitrary choice of sampling protocol as if it were physical information. The latent transition of Eq.~\eqref{eq:latent_transition} avoids this by construction. What the network \(f_\theta\) learns is not an increment but a rate, \(d\z/d\eeq{}\) in the continuum limit of Eq.~\eqref{eq:latent_ode} with \(\kappa=\eeq\), defined with respect to equivalent-strain progression rather than time. The actual size of the latent-state update at a given step is then obtained only afterward, by multiplying this rate by the measured equivalent-strain increment \(|\Delta\eeqt_t|\) between consecutive observations. Two recordings of the same physical loading path sampled at different densities therefore differ only in how many times the same underlying rate law is queried and integrated, not in the rate law itself, so the learned constitutive response is time-resolution-independent and can be built purely from heterogeneous, unevenly sampled experimental observations without first resampling them onto a common time base. 

This rate-based construction further justifies a temporal downsampling to reduce data demands during training. Since the transition model already treats sampling density as a numerical, not a physical, degree of freedom, reducing the number of retained points along a trajectory only coarsens the numerical integration of Eq.~\eqref{eq:latent_ode} without altering the underlying rate being learned.

\begin{figure}[!ht]
    \centering
    \includegraphics[width=\textwidth]{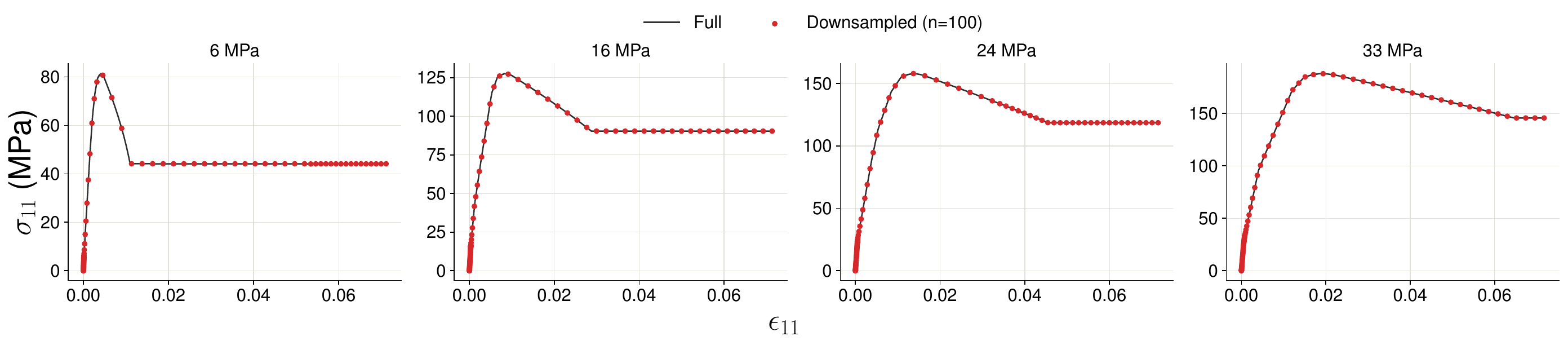}
    \caption{Effect of temporal downsampling on representative axial stress--strain training sequences, shown as \(\sigma_{11}\) against \(\epsilon_{11}\) at several confining pressures. Black lines denote the full-resolution sequence after plateau truncation; red markers denote the \(N_{\mathrm{tpts}}=100\) points retained by evenly spaced index selection. The elastic branch, peak, and post-peak plateau are all preserved in the reduced sequence.
    }
    \label{fig:downsample}
\end{figure}

A full-resolution triaxial history typically has more time steps than the recurrent latent-state model needs, particularly along the post-peak plateau, where stress stays nearly constant over a long strain range. Each sequence is therefore subsampled along its loading axis to a fixed number of points, after the post-peak plateau truncation and normalization-statistics computation of Section~\ref{sec:norm} have already been applied to the full-resolution data. A curvature-based scheme that concentrates points near the pre-peak rise and peak could prioritize coverage of the mechanically decisive regions. However, this work uses the simpler alternative of evenly spaced indices along the truncated sequence, which retains the pre-peak rise, peak, and post-peak plateau in the same proportion as the original record. Every per-time-step channel (strain, stress, equivalent strain, and the invariants \(I_1\) and \(J_2\) together with their lagged counterparts) is indexed identically, preserving temporal alignment across channels exactly. Figure~\ref{fig:downsample} shows that the retained points track the full-resolution curve through the elastic branch, peak, and residual plateau without visible loss of resolution.

Two downsampling targets are used during optimization, reflecting the different roles of the training and validation calculations. Training sequences, and the corresponding non-autoregressive evaluation, are reduced to \(N_{\mathrm{tpts}}=100\) points per sequence, while the more computationally expensive autoregressive validation rollout used for checkpoint selection is evaluated on \(N_{\mathrm{vpts}}=500\) points per sequence. This optimization-time validation is distinct from the final offline evaluation in Section~\ref{sec:case1} where, after checkpoint selection, both validation and test sequences are evaluated over their complete preprocessed time histories, without temporal downsampling. Downsampling is performed once during dataset construction rather than being redrawn at each training epoch, so that all epochs train on the same fixed set of points per sequence.

\subsection{Rotational augmentation}

Here the material is treated as isotropic at the scale of the representative volume element (RVE), so that the constitutive response depends on the current strain state and loading history alone, not on the orientation of that strain state relative to any fixed material texture. This is consistent with classical continuum and microplane treatments of plain concrete and related quasi-brittle materials, which idealize the initial material state as statistically isotropic and introduce no prescribed material texture or preferred orientation \citep{bazant1985microplane,bazant1998fracture,lubliner1989plasticdamage}. The assumption matters in practice because the available triaxial histories are all recorded in a single laboratory frame, with the axial loading direction aligned to a fixed coordinate axis, while a finite element mesh built from this model (in particular the multi-element structural model) presents strain states at whatever orientation the local element axes happen to have relative to that frame. A stress map trained only on axis-aligned data has no guarantee of behaving consistently once queried far from that orientation, because frame invariance is not automatically guaranteed in neural-network constitutive models and must be enforced through suitable inputs, architectures, loss functions, or data augmentation \citep{heider2020so3,vlassis2020geometric}.

Under RVE-scale isotropy, the constitutive law must satisfy the classical frame-indifference condition \citep{truesdell2004nonlinear}. For a proper rotation \(\mathbf{R}\in\mathrm{SO}(3)\), strain and stress tensors transform as \(\eps'=\mathbf{R}\eps\mathbf{R}^\top\) and \(\sig'=\mathbf{R}\sig\mathbf{R}^\top\), and the stress map must satisfy the equivariance condition \(g(\mathbf{R}\eps\mathbf{R}^\top,\cdot)=\mathbf{R}\,g(\eps,\cdot)\,\mathbf{R}^\top\) for every \(\mathbf{R}\). The sparse GP of Section~\ref{sec:formulation} is a generic nonparametric regressor with no architectural mechanism that enforces this condition exactly, unlike equivariant network constructions built specifically for this purpose \citep{heider2020so3,vlassis2020geometric}. Rotational augmentation instead aims to satisfy this constraint in an approximate, statistical form. By training on many rotated copies of the same physical loading history, the model is encouraged to interpolate consistently across orientations that its architecture does not constrain directly.

Concretely, a fixed set of \(N_R=20\) rotations is constructed once and applied to every retained
stress--strain history. For each history, the same \(\mathbf{R}\) is applied to every strain and stress tensor
pair \((\eps_t,\sig_t)\) along a given history. The set is a deterministic quasi-uniform covering
of \(\mathrm{SO}(3)\) rather than a finite random sample. Five axes are placed on \(S^2\) by the
golden-angle construction, and each axis is combined with four equally spaced in-plane rotations
about it, giving \(5\times4=20\) orientations.
It is essential that \(\mathbf{R}\) be held fixed along a given trajectory rather than resampled at each time step. A time-varying rotation would not correspond to a change of orientation at all, but would superimpose a spurious, time-varying rigid rotation on top of the physical deformation, corrupting the sequential structure from which the latent transition of Section~\ref{sec:formulation} is meant to learn. The \(N_R=20\) rotated copies of each sequence are added to the training set only. Validation and
test sequences, and the normalization statistics of Section~\ref{sec:norm}, are computed from
unrotated data, so that checkpoint selection and all reported errors are measured on physically
distinct loading histories rather than on rotations of sequences the model has already seen. The
training set therefore comprises \(N_{\text{train}}\times(1+20)\) sequences,
and the six-component GP stress map is exposed to many orientations of the same underlying
material response.

The latent transition and the stress map require different treatments under rotation. The transition updates the latent state using only the scalar descriptors \(\eeq\), \(I_1\), and \(J_2\), all of which are unchanged when the strain and stress tensors are rotated by \(\mathbf{R}\). Consequently, rotating an entire loading history leaves the inputs to the transition unchanged and produces the same latent-state trajectory. That is, rotational augmentation adds no new information to this part of the model. The stress map \(g\), by contrast, receives the components of the strain tensor and returns the components of the stress tensor. It must therefore satisfy the equivariance relation
\begin{equation}
g(\mathbf{R}\eps\mathbf{R}^{\top},\z)
=\mathbf{R}\,g(\eps,\z)\,\mathbf{R}^{\top},
\end{equation} 
which is not enforced by the generic sparse-GP architecture. The rotated stress--strain pairs are introduced specifically to encourage this relation during training. Thus, rotational consistency is exact for the invariant-driven latent transition but approximate and learned from augmented data for the tensorial stress map.

It's important to emphasize that the assumptions in this section would, of course, break down for a material with a fixed anisotropic microstructure, such as bedding planes, joint sets, or a preferred fabric orientation. In such cases, testing along a rotated axis would probe a genuinely different response that could not be recovered by rotating the original data. 

\subsection{Training protocol}
\label{sec:training}

Training is performed by minimizing the negative ELBO of Eq.~\eqref{eq:svgp_elbo}
for the sparse multi-output GP together with the neural transition parameters, in the normalized form of Eq.~\eqref{eq:minibatch_elbo}, using a small number of cross-validation folds trained with different random seeds. Each fold uses a fixed, moderate number of inducing points, a bounded number of training epochs with early stopping on validation performance, a fixed learning rate, and the \(S\)-sample Monte Carlo average over stochastic latent state rollouts defined in Eq.~\eqref{eq:total_training_loss}, evaluated on the downsampled training and validation sequences of Section~\ref{sec:downsample}. The expected log-likelihood inside the ELBO is
evaluated in the closed form given in Eq.~\eqref{eqn:ELBO_loglike}. Exact hyperparameter values are reported in \ref{sec:implementation}.

The best checkpoint is selected by autoregressive validation RMSE in normalized stress units. The corresponding optimization-time RMSE and epoch are reported in \ref{sec:implementation}. After checkpoint selection, the model is evaluated offline on the complete validation and test time histories in Section~\ref{sec:case1} before being exported for the finite element deployment cases in Sections~\ref{sec:case2} -- \ref{sec:case4}.

In the next section, we first verify the learned constitutive relation against complete RVE-scale histories. We then test whether that relation can be exported to and correctly executed within a commercial finite element package (in this case Abaqus), from a single integration point to a multi-element structural simulation. Our objective is therefore verification of the learned material-point update and its implementation and deployment pathway in Abaqus/Explicit.

\section{Results}
\label{sec:results}

We proceed through four cases of increasing complexity. Case~1 (Section~\ref{sec:case1}) evaluates the trained GP-SSM on complete validation and test histories and therefore establishes the constitutive law prediction to be deployed, independently of its finite element implementation. Case~2 (Section~\ref{sec:case2}) then documents the full Abaqus/Explicit deployment pathway, from the exported parameter file to the VUMAT update loop, and exercises it under a triaxial path in which the confining stress changes during loading. Case~3 (Section~\ref{sec:case3}) extends the deployed material-point model to include an unloading--reloading rule that freezes the learned latent memory whenever the equivalent strain drops below its historical maximum, an ingredient required before structural simulation because some material points may unload while others continue to load to the softening regime. Case~4 (Section~\ref{sec:case4}) uses that rule in a multi-element structural model that reproduces a loading condition from the KCC literature and compares the resulting structural compression response against the KCC model as a reference solution. 

In all cases, the GP-SSM is developed using \(N_{\mathrm{hist}}=20\) synthetic triaxial compression histories generated with the KCC model implemented as an Abaqus/Explicit VUMAT. The data-generation simulations follow the loading conditions used in Figure~4 of \citet{wu2015numerical}, with monotonic axial compression applied under fixed lateral confinement. The dataset spans confining pressures from \(2\) to \(40~\mathrm{MPa}\), with each history recording the response to monotonic axial compression under fixed lateral confinement. These histories capture the initial loading branch, peak strength, and post-peak softening toward residual response, covering the transition from a sharp stress drop at low confinement to more gradual softening and higher sustained strength at elevated confinement. The data are partitioned by complete loading history: twelve sequences at \(2, 6, 8, 12, 16, 20, 22, 24, 30, 33, 36,\) and \(38~\mathrm{MPa}\) are used for training, five at \(10, 14, 18, 31,\) and \(40~\mathrm{MPa}\) for validation, and three at \(4, 26,\) and \(28~\mathrm{MPa}\) for testing. The model learns from the strain and stress histories, with equivalent strain and stress invariants calculated from these quantities; KCC internal variables are not supplied as training targets. Each training sequence is reduced to 100 temporally aligned points after preprocessing, whereas the final validation and test evaluations use the complete preprocessed histories.

\begin{figure}[!ht]
    \centering
    \includegraphics[width=\textwidth, trim= 0.5cm 9cm 4cm 4cm,clip]{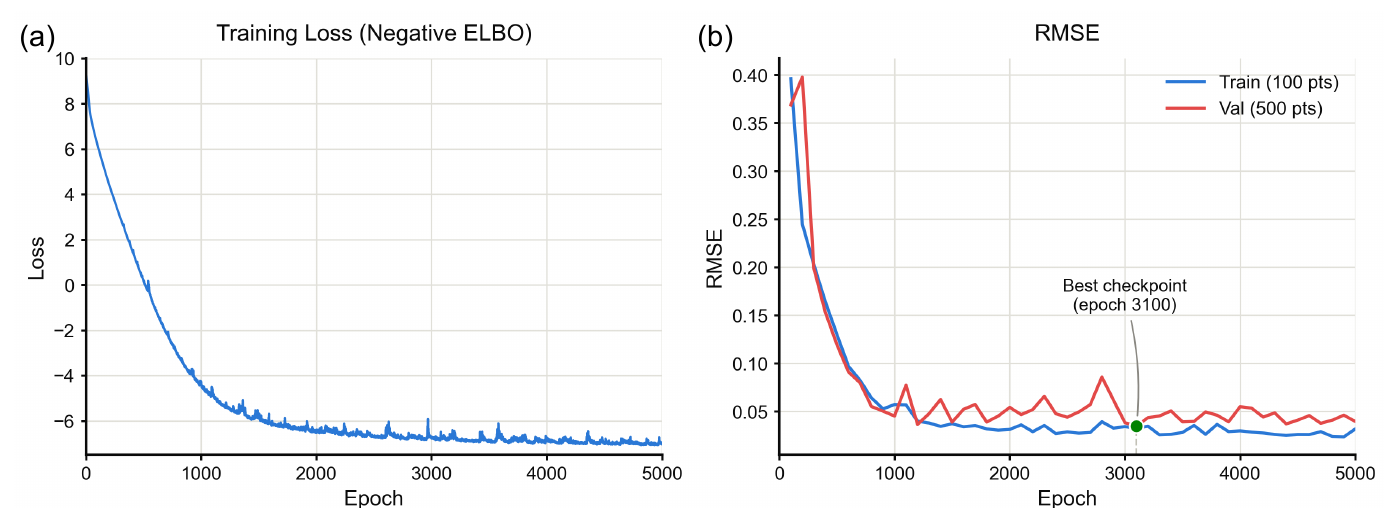}
    \caption{Training dynamics of the GP-SSM (representative fold): (a) Negative-ELBO training loss versus epoch; (b) Train and validation RMSE evaluated every 100 epochs, where train RMSE is computed non-autoregressively on $N_{\mathrm{tpts}} = 100$ downsampled points per sequence and validation RMSE uses closed-loop autoregressive rollout on $N_{\mathrm{vpts}} = 500$ points per sequence, with predicted stress fed back as the invariants at each step. The marked point (epoch 3100) indicates the checkpoint selected by early stopping on the EMA-smoothed validation RMSE.}
    \label{fig:training_curves_pub}
\end{figure}

Figure~\ref{fig:training_curves_pub} summarizes the training dynamics for a representative fold. The negative-ELBO loss in Figure~\ref{fig:training_curves_pub}(a) decreases rapidly during the initial epochs and then improves more gradually, approaching a plateau with small fluctuations. Figure~\ref{fig:training_curves_pub}(b) shows a corresponding reduction in both training and validation RMSE. The validation error fluctuates more strongly because it is evaluated through closed-loop autoregressive rollout, whereas the training error uses reference previous-step stress invariants. After the initial improvement, further reductions in training loss yield limited gains in validation accuracy. The checkpoint at epoch 3100 is therefore selected using the exponentially moving average of the validation RMSE, with a normalized validation RMSE of approximately \(0.0347\), and is used for the subsequent evaluations and finite element deployment.

\subsection{Case 1: Offline full-sequence rollout evaluation}
\label{sec:case1}

\begin{figure}[!ht]
    \centering
    \includegraphics[width=\textwidth]{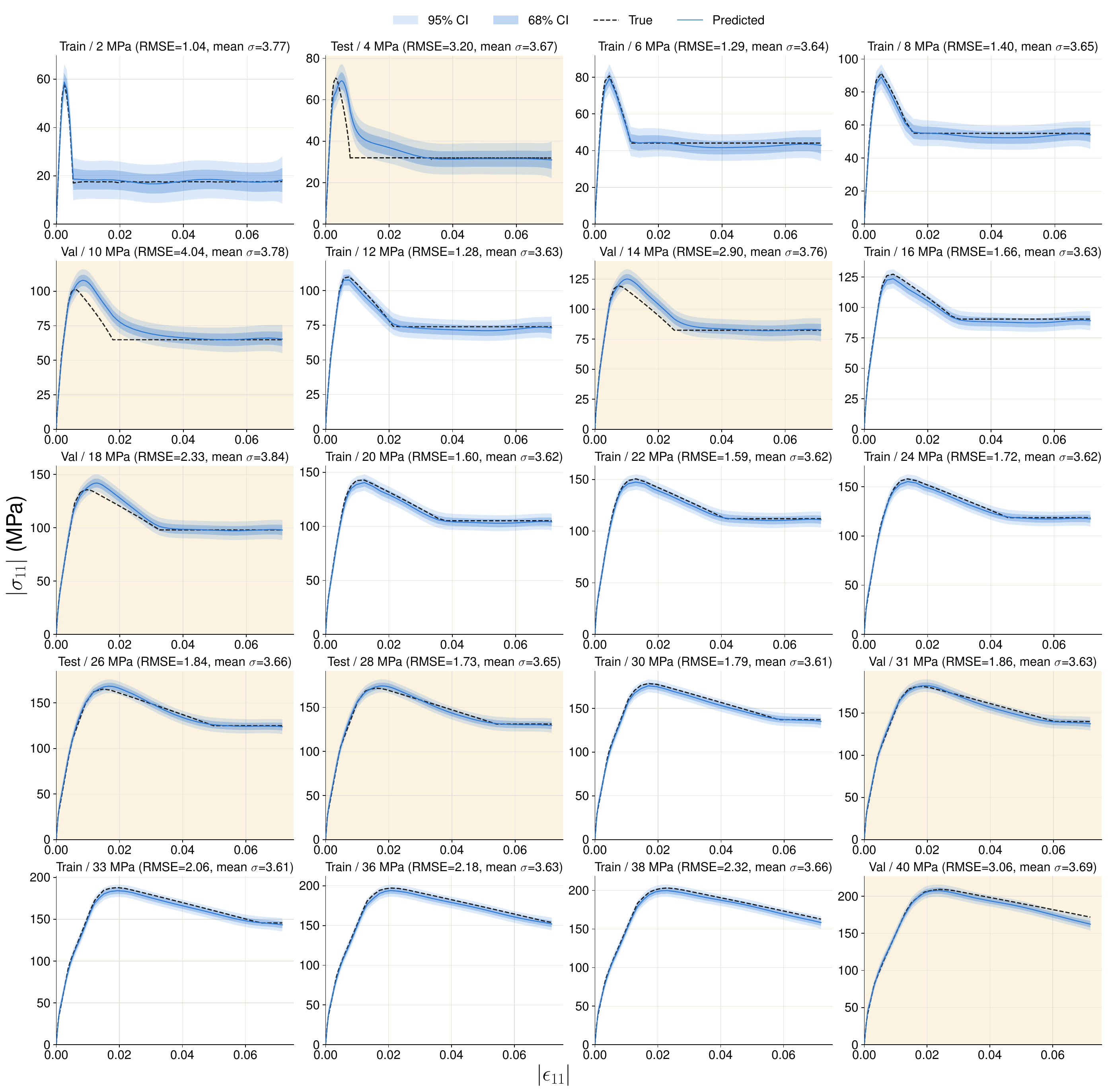}
    \caption{Offline GP-SSM predictions of the axial stress--strain response under triaxial compression with reported levels of confinement, shown as $|\sigma_{11}|$ against $|\epsilon_{11}|$ for each confining pressure (labeled in MPa). Black dashed lines denote the reference RVE-scale stress histories. Solid blue lines denote the GP-SSM posterior mean, and the shaded bands show the 68\% and 95\% credible intervals. Training sequences (white background) are evaluated non-autoregressively using the reference previous-step stress invariants. Validation and test sequences (yellow background) are evaluated in closed-loop autoregressive rollout over their complete preprocessed time histories, without temporal downsampling, with predicted stress fed back to compute the invariants at the next step. RMSE and mean predictive standard deviation are reported for each sequence.}
    \label{fig:offline_rollouts}
\end{figure}

Case~1 is performed offline with respect to the finite element package. After the checkpoint is selected using the optimization-time validation procedure of Section~\ref{sec:training}, the model is evaluated on the complete preprocessed time history of every sequence shown in Figure~\ref{fig:offline_rollouts} (the black dashed lines in the figure denote the reference RVE-scale stress histories). Neither the validation sequences nor the test sequences (yellow background in Figure~\ref{fig:offline_rollouts}) are temporally downsampled for this final evaluation. At each time step, the measured strain is supplied to the model, the latent state is advanced incrementally, and the predicted stress, rather than the reference stress, is used to compute the invariants supplied to the next step. The validation and test results are therefore closed-loop autoregressive rollouts over the full time series, and assess accumulated sequential prediction error rather than isolated one-step accuracy.

As shown in Figure~\ref{fig:offline_rollouts}, the model reproduces the systematic increase in peak and residual axial strength with confinement across the full set of RVE-scale histories. For the training sequences, the RMSE ranges from \(1.04\) to \(2.32~\mathrm{MPa}\). More importantly for offline generalization, the full-sequence validation RMSE ranges from \(1.86\) to \(4.04~\mathrm{MPa}\), and the test RMSE ranges from \(1.73\) to \(3.20~\mathrm{MPa}\). The closest agreement occurs for the more highly confined test histories at \(26\) and \(28~\mathrm{MPa}\), where the predicted curves track the reference loading, peak, softening, and residual response. The largest discrepancies occur at low confinement, particularly for the \(10~\mathrm{MPa}\) validation history and the \(4~\mathrm{MPa}\) test history, where the reference response exhibits a sharp post-peak stress drop that is represented by a smoother transition in the GP-SSM prediction. The predictive intervals also widen in these low-confinement cases, consistent with their greater rollout error. These results establish that the trained state-space update remains stable when rolled out over every point of a held-out sequence and interpolates across unseen confinement levels within the available triaxial loading family. They do not, by themselves, establish generalization to loading paths outside that family. Instead, Cases~2--4 test whether this offline constitutive update can be implemented and exercised correctly in finite element analyses.

\subsection{Deployment within finite element software}
\label{sec:case2}

\begin{figure}[!ht]
    \centering
    \includegraphics[width=0.8\textwidth,trim=5cm 3.2cm 3.5cm 4.2cm,clip]{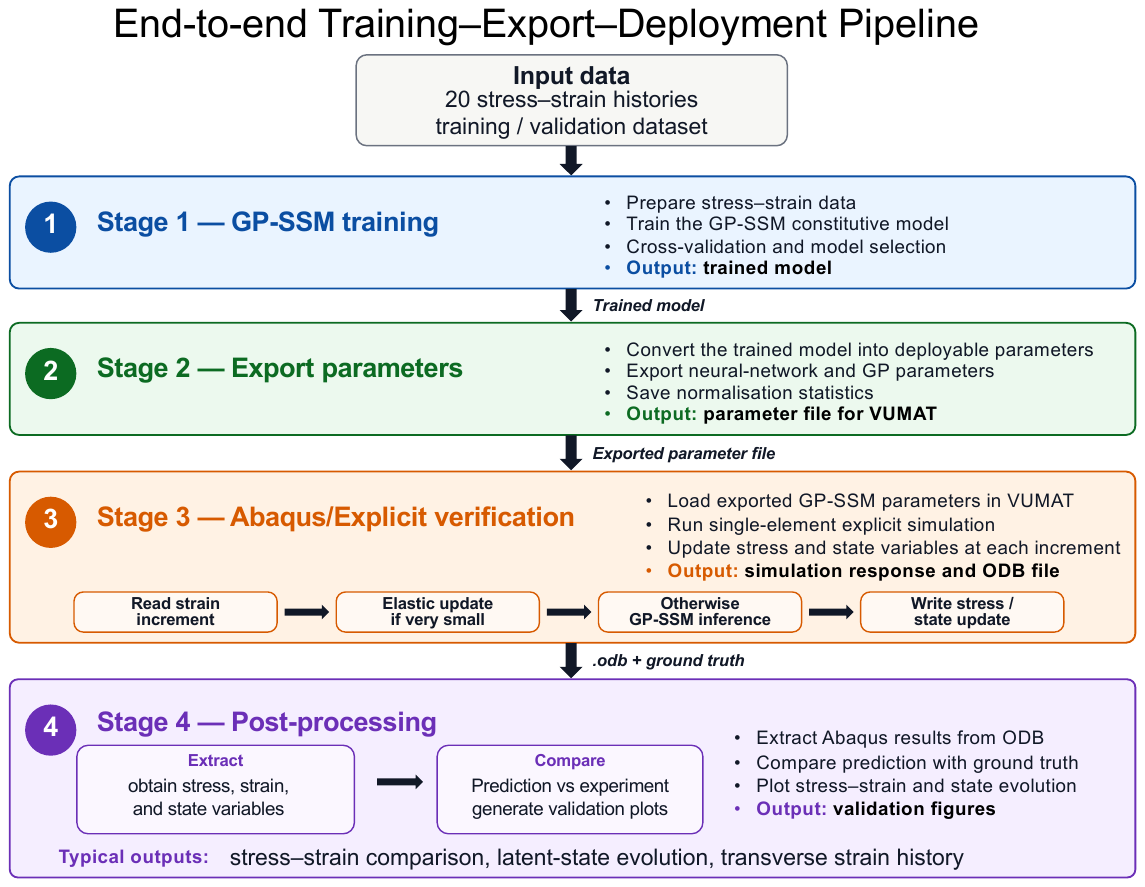}
    \caption{End-to-end pipeline from experimental stress--strain data and GP-SSM training to parameter export, VUMAT implementation, and finite element deployment in Abaqus/Explicit.}
    \label{fig:pipeline}
\end{figure}

Here, in Case 2, we make the complete pipeline explicit, from experimental data to a verified finite element deployment, shown in Figure~\ref{fig:pipeline}. Training uses only strain and stress histories that are directly measurable in a standard triaxial compression test, without measuring or prescribing damage, plastic strain, or any other internal variable. The latent state \(\z_t\) that plays this role is instead inferred statistically, through the variational training procedure of Section~\ref{sec:formulation}. The same inferred history representation, not a hand-specified damage or plasticity variable, advances the material state once the trained model is embedded in the finite element solver (in this case, Abaqus/Explicit). This experiment-to-deployment pipeline is exercised end-to-end for the first time in this Case~2.

\subsubsection{Deployment protocol}
After training, the model parameters are exported through a lightweight parameter file representation. The exported parameter file contains the four affine layers of the transition network, the inducing-point matrix, ARD length scales, output scales, GP mean constants, posterior-mean coefficients, and normalization statistics. In the present model, the plain-text representation contains 10,994 scalar parameters.
The exported representation is evaluated inside a VUMAT user material subroutine in Abaqus/Explicit, and Cases~2--4 execute through this same deployment pathway.
This file-based representation is intentionally simple. The VUMAT reads the parameters once, stores them in local arrays, and performs only elementary floating-point operations during the material-point update.

At each material point and time increment, the subroutine performs the following local update:
\begin{enumerate}[leftmargin=2em]
    \item Update the accumulated strain from the strain increment supplied by the explicit solver;
    \item Compute \(\eeq\), \(I_1\), and \(J_2\) from the current or previous material state;
    \item Update the latent state using Eq.~\eqref{eq:latent_transition};
    \item Evaluate the GP posterior mean using Eq.~\eqref{eq:gp_mean};
    \item Transform the normalized stress back to physical units and store the updated latent variables and invariants as state variables.
\end{enumerate}
The VUMAT state variables are summarized in Table~\ref{tab:sdv}. The two latent variables are followed by the six accumulated strain components and the previous-step stress invariants. This keeps the update local at each integration point and avoids global history storage.

\begin{table}[!ht]
\centering
\caption{State variables used by the Abaqus/Explicit VUMAT implementation.}
\label{tab:sdv}
\begin{tabular}{cl}
\toprule
State variable index & Meaning \\
\midrule
1--2 & latent state components \(z_1,z_2\) \\
3--8 & accumulated six-component strain vector \\
9 & previous first stress invariant \(I_1\) \\
10 & previous deviatoric invariant \(J_2\) \\
\bottomrule
\end{tabular}
\end{table}

Because Abaqus/Explicit supplies logarithmic strain increments to VUMAT, the accumulated normal logarithmic strains are converted to engineering normal strains by \(\varepsilon_i=\exp(\varepsilon_i^{\log})-1\) before evaluating the GP-SSM. The shear components remain small in the present single-element verification and are passed through without additional nonlinear conversion. For very small equivalent strain, specifically \(\eeq<5\times10^{-4}\), a linear elastic response is used as an initialization branch because the training data do not densely cover the near-zero strain regime. The elastic branch uses \(K=16370~\mathrm{MPa}\), \(G=12790~\mathrm{MPa}\), and \(\kappa=K-2G/3\), corresponding to a Young's modulus \(E=30.4~\mathrm{GPa}\) and Poisson's ratio \(\nu=0.188\). Note that these specific material parameters are taken directly from the KCC model. 
A conservative extrapolation guard is also used outside the training-supported equivalent-strain range by freezing the stress and latent state when \(\eeq>\mu_{\eeq}+4s_{\eeq}\).
Additionally, if the accumulated shear strain is essentially zero, the corresponding GP shear-stress output is suppressed to prevent small numerical shear noise from being amplified by unconstrained shear accelerations. These safeguards are numerical devices for stable finite element deployment and are not part of the conceptual GP-SSM formulation.

\subsubsection{Case 2: Single-element compression under changing confinement}
\begin{figure}[!ht]
    \centering
    \includegraphics[width=\textwidth,
    trim=1cm 4cm 7cm 3cm,
    clip]{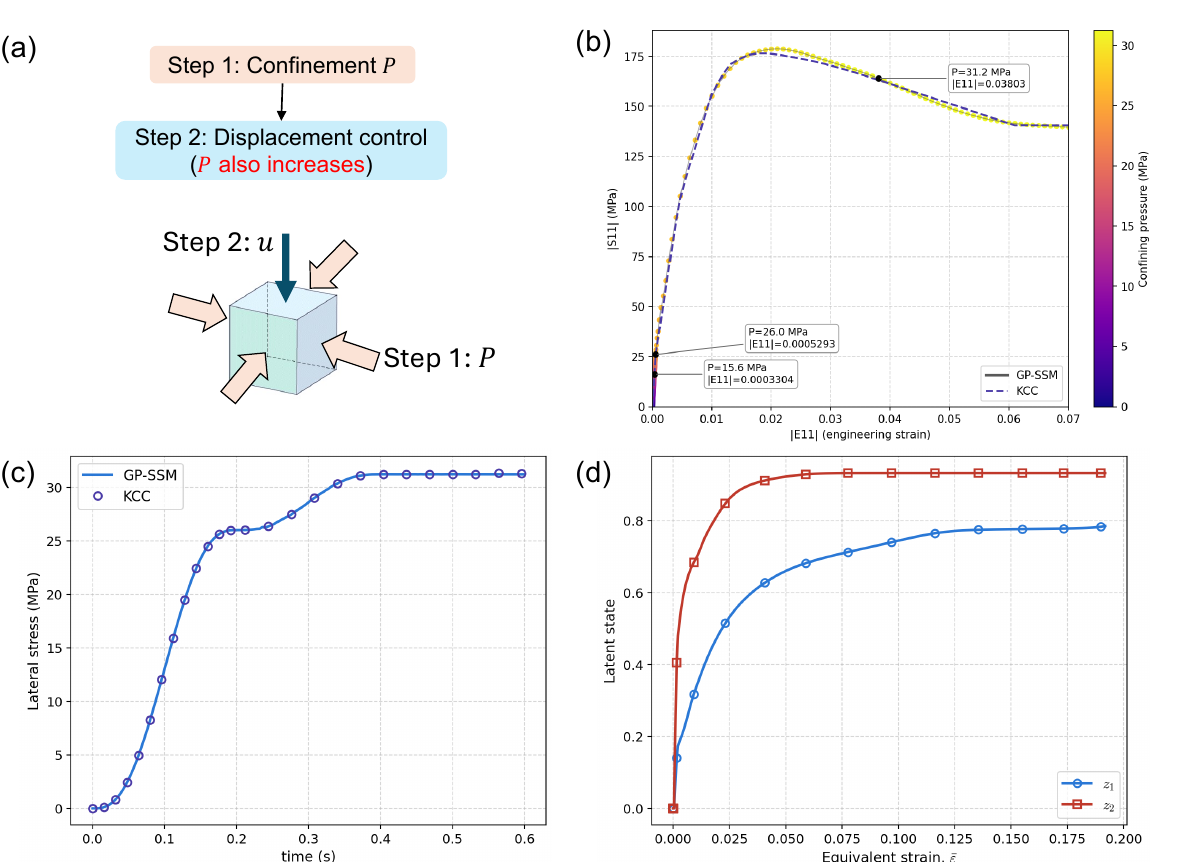}
    \caption{Changing-confinement loading path and response: (a) the two-step loading protocol, with confining pressure \(P\) ramped from \(0\) to \(26~\mathrm{MPa}\) in Step~1 and continuing to rise smoothly from \(26\) to \(31.2~\mathrm{MPa}\) during the axial displacement/velocity loading of Step~2. (b) axial stress--strain response \(|S_{11}|\) versus \(|E_{11}|\), colored by the instantaneous confining pressure, comparing the GP-SSM model against the KCC reference model, with confining-pressure values annotated at representative points along the curve. (c) lateral stress evolution versus time, showing close agreement between the GP-SSM and the reference KCC model as the confining pressure rises through both loading steps. (d) evolution of the latent state components \(z_1\) and \(z_2\) with equivalent strain \(\bar\varepsilon\), showing monotonic saturation consistent with accumulating material history.
    }
    \label{fig:case2_changing_confinement}
\end{figure}

Case~2 examines one reduced-integration eight-node brick element with 101.6 mm side length under a two-step triaxial loading path, with minimal additional velocity constraints to remove rigid-body motion. In Step~1, a confining pressure \(P\) is smoothly ramped from \(0\) to \(26~\mathrm{MPa}\) on the lateral element faces, establishing a hydrostatic compression state. In Step~2, an axial displacement/velocity boundary condition is smoothly applied to impose compressive axial loading, while the confining pressure continues to increase from \(26\) to \(31.2~\mathrm{MPa}\). This is illustrated in Figure~\ref{fig:case2_changing_confinement}(a). All of these specific pressures are chosen to retain consistency with existing studies with the KCC model. This non-constant-confinement path directly exercises the previous-step invariants \(I_1\) and \(J_2\) of Eq.~\eqref{eq:qb_assignment}, since the pressure driving the latent transition evolves throughout axial loading.

As shown in Figure~\ref{fig:case2_changing_confinement}(c), the lateral stresses \(S_{22}\) and \(S_{33}\) closely track the imposed confining pressure through both steps. They follow the smooth ramp to \(26~\mathrm{MPa}\) during Step~1 and then continue to track the further rise to \(31.2~\mathrm{MPa}\) during Step~2, rather than remaining pinned at the Step-1 level. The axial response \(|S_{11}|\)  shown in Figure~\ref{fig:case2_changing_confinement}(b) rises to a peak of approximately \(178~\mathrm{MPa}\) at an axial strain of about \(0.02\), then softens to a plateau of approximately \(140~\mathrm{MPa}\) by an axial strain of about \(0.06\)--\(0.07\). The finite element model implementing the GP-SSM constitutive law through VUMAT remains in close agreement with the KCC reference model throughout, including through the peak and softening regions, confirming that the finite element deployment reproduces the underlying material-point model correctly under this non-constant-confinement path. Because the confining pressures at the annotated points in Figure~\ref{fig:case2_changing_confinement}(c)
fall within the range of the training data, the results indicate that the material response is capable of evolving with the changing confining pressure rather than being governed solely by its initial value. Note that this type of boundary condition was not included in the training process, demonstrating the model's ability to generalize to an unseen loading history.
Figure~\ref{fig:case2_changing_confinement}(d) shows that both latent components increase monotonically with equivalent strain, with \(z_2\) saturating earlier than \(z_1\). This behavior reflects the nonnegative transition constraint: the latent state retains accumulated loading history even during post-peak stress softening, while its evolution gradually slows with further deformation.

\subsection{Case 3: Unloading--reloading extension}
\label{sec:case3}

The RVE-scale histories evaluated in Case~1 and the finite element loading path in Case~2 are monotonic: equivalent strain never decreases, and the sign constraint of Eq.~\eqref{eq:qb_assignment} means the latent state only accumulates. A structural simulation cannot be restricted to monotonic loading at every material point. Once some region of a structure begins to soften, deformation may localize into a narrow band while neighboring material points unload. For local strain-softening constitutive laws, this localization can lead to loss of well-posedness and mesh sensitivity. A substantial literature addresses this issue by introducing an internal length scale through crack-band models \citep{bazant1983crackband}, integral nonlocal damage \citep{pijaudier1987nonlocal,bazant2002nonlocal}, gradient-enhanced damage \citep{peerlings1996gradient}, phase-field fracture models \citep{miehe2010phasefield}, and (most recently) a smooth crack band model (sCBM) using ``sprain energy," which introduces the curvature of the displacement field as an explicit localization limiter \citep{xu2024sprain}. Reproducing these methods is not the objective of the present paper. Instead, Case~3 introduces a physically consistent elastic unloading branch for the material-point update, so that individual integration points can unload without violating the irreversibility built into the learned latent transition. This supplies the minimum ingredient needed to perform multi-element structural simulation.

This unloading branch is built on the equivalent-strain memory already used to gate the extrapolation guard in Section~\ref{sec:case2}. Let
\begin{equation}
\eeqmax_t=\max\left(\eeqmax_{t-1},\,\eeq_t\right)
\end{equation}
denote the running maximum equivalent strain reached up to increment \(t\). Let \(t_f\) 
denote the increment at which the current unloading excursion begins. The latent state, stress, strain, and stress invariants stored at that increment are denoted by \(\z_{t_f}\), \(\sig_{t_f}\), \(\eps_{t_f}\), \(I_{1,t_f}\), and \(J_{2,t_f}\), respectively. If the current equivalent strain exceeds the prior maximum, \(\eeq_t=\eeqmax_t>\eeqmax_{t-1}\), the material point follows the ordinary GP-SSM update of Eq.~\eqref{eq:latent_transition}. If instead \(\eeq_t<\eeqmax_{t-1}\), indicating unloading or reloading below the previous maximum, the latent state and invariant memory are held at the values stored at \(t_f\),
\begin{equation}
\z_t=\z_{t_f},
\qquad
\eeqmax_t=\eeqmax_{t-1},
\end{equation}
and the stress is instead updated from the frozen reference state through the linear elastic stiffness,
\begin{equation}
\sig_t=\sig_{t_f}+\mathbf{C}^{e}:\left(\eps_t-\eps_{t_f}\right).
\label{eq:elastic_unloading}
\end{equation}
If at a later time the equivalent strain again exceeds \(\eeqmax\), the material point resumes the learned GP-SSM update of Eq.~\eqref{eq:latent_transition} from the frozen latent state \(\z_{t_f}\), rather than restarting from the current elastic stress. This construction is deliberately minimal. It does not introduce a new learned unloading stiffness or a hysteresis model, but it prevents the irreversible, monotonically evolving latent state from being pulled backward by elastic unloading, which would be inconsistent with its interpretation as an accumulated history variable. 
For repeated excursions, \(t_f\) is updated at the onset of each new unloading segment. 
Note that \(\mathbf{C}^{e}\) here is the initial, undamaged stiffness and does not depend on \(\z_{t_f}\). Unloading occurs at the initial modulus regardless of accumulated history, with no stiffness degradation and no hysteretic dissipation. Case~3 tests only the irreversibility of the latent state, not a calibrated cyclic response.  

More generally, cyclic loading with unloading--reloading stiffness degradation and hysteresis can also be accommodated within the GP-SSM framework by conditioning a learned unloading--reloading rule on the frozen latent state. Because \(\z_{t_f}\) retains learned information accumulated over the preceding loading history, it can serve the same functional role as classical internal or history variables used to represent damage, plastic strain, and related irreversible processes. This interpretation does not require assigning any individual latent coordinate a one-to-one physical meaning. Rather, the latent state collectively provides the history dependence needed to learn degraded stiffness and hysteretic response. These extensions will be considered in future work.

\begin{figure}[!ht]
    \centering
    \includegraphics[width=\textwidth,
    trim=1cm 3cm 7cm 3cm,
    clip]{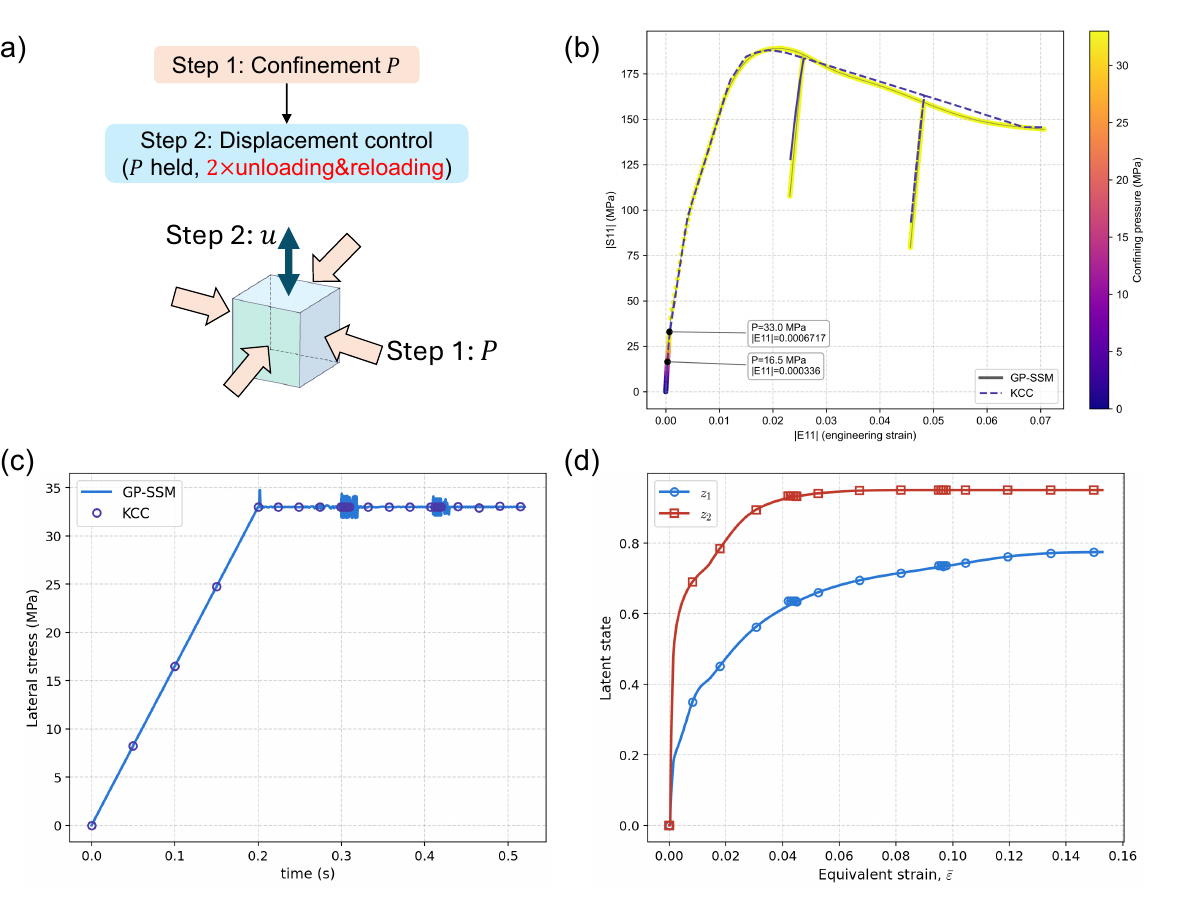}
    \caption{Single-element response under fixed confinement with two unloading--reloading excursions: (a) The two-step protocol, with confining pressure \(P\) ramped from \(0\) to \(33~\mathrm{MPa}\) in Step~1 and held constant in Step~2 while an axial displacement \(u\) is imposed and twice reversed. (b) Axial response, \(|S_{11}|\) against \(|E_{11}|\) showing two elastic unloading events. The solid line is the GP-SSM prediction colored by current confining pressure. The dashed line is the KCC reference under the same load path. (c) Lateral stress versus time, showing the ramp to \(33~\mathrm{MPa}\) and subsequent maintenance of confinement, with brief fluctuations during unloading–reloading. (d) Latent components \(z_1\) and \(z_2\) versus equivalent strain, remaining fixed during unloading–reloading below the historical maximum and resuming evolution once that maximum is exceeded.
    }
    \label{fig:loading_unloading}
\end{figure}

Case~3 reuses the single-element setup of Case~2 with a fixed-confinement loading path. As illustrated in Figure~\ref{fig:loading_unloading}(a), the confining pressure is first ramped to \(33~\mathrm{MPa}\) and then held constant while axial compression is applied through a prescribed velocity boundary condition on the loaded face. Two unloading--reloading excursions are introduced by reversing the axial velocity to unload the element and then reversing it again to resume compression beyond the previous maximum equivalent strain. Figure~\ref{fig:loading_unloading}(b) shows the resulting axial stress--strain response, with linear elastic unloading and reloading branches governed by \(\mathbf{C}^e\) about the frozen reference state, as defined in Eq.~\eqref{eq:elastic_unloading}. Throughout these excursions, the lateral stress remains close to the prescribed confinement, apart from brief fluctuations associated with the loading reversals, shown in  Figure~\ref{fig:loading_unloading}(c). During unloading and subsequent reloading below the historical maximum equivalent strain \(\eeqmax\), the frozen-memory rule holds both the latent state and the stored stress invariants fixed. Once the equivalent strain exceeds \(\eeqmax\), the ordinary GP-SSM update of Eq.~\eqref{eq:latent_transition} resumes from the latent state stored at \(t_f\). Figure~\ref{fig:loading_unloading}(d) highlights that both latent components remain unchanged throughout unloading and reloading below \(\eeqmax\), confirming that these excursions preserve the accumulated material memory.

\subsection{Case 4: Structural application}
\label{sec:case4}

\begin{figure}[!ht]
    \centering
    \includegraphics[width=\textwidth,
    trim=3.5cm 3.5cm 4cm 3cm,
    clip]{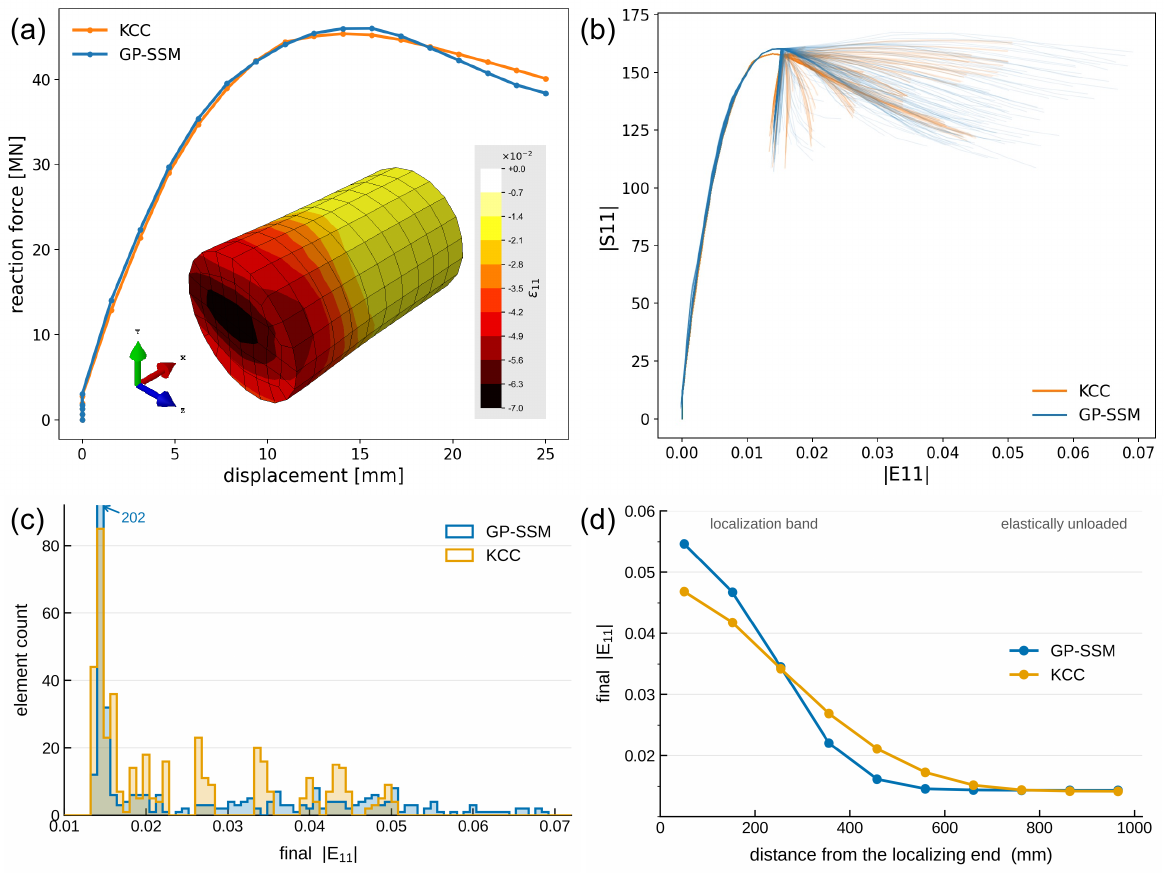}
    \caption{Structural-level finite element response of the multi-element cylindrical model under \(24~\mathrm{MPa}\) confinement, comparing the GP-SSM-driven simulation against the KCC reference solution. (a) axial reaction force versus imposed end displacement, with an inset showing the deformed mesh. (b) element-level axial stress--strain trajectories for both models, illustrating the heterogeneous loading and unloading histories that develop across the cylinder. (c) histogram of the final axial strain \(|E_{11}|\) at \(25~\mathrm{mm}\) end displacement over all elements for the GP-SSM (blue) and KCC (orange) simulations. (d) layer-averaged final axial strain \(|E_{11}|\) as a function of axial distance from the localizing end, showing that both models concentrate the deformation within roughly the first \(400~\mathrm{mm}\) of the cylinder and recover the same elastically unloaded plateau beyond it, with the GP-SSM predicting a slightly more intense and more sharply terminated localization band.}

    \label{fig:case4_structural}
\end{figure}

Case~4 applies the GP-SSM constitutive update to a cylindrical structural model with diameter \(609.6~\mathrm{mm}\) and length \(1016~\mathrm{mm}\), with its axis aligned with the global \(x\)-direction. The model is discretized with 430 eight-node reduced-integration brick elements (C3D8R) and 594 nodes as illustrated in Figure~\ref{fig:case4_structural}(a).
The end face at \(x=1016~\mathrm{mm}\) is fixed in the axial direction, with its center node additionally fixed in the two transverse directions. The $x=0$ face is displaced in the $x$ direction and are unconstrained in the transverse ($y$ and $z$) directions except for 
two longitudinal node lines, which are constrained to remove the remaining rigid-body modes. 
A uniform pressure is applied to the cylindrical surface and ramped linearly from zero to \(24~\mathrm{MPa}\) over the first \(0.2~\mathrm{s}\), after which it is held constant. The axial displacement remains zero during this confinement ramp and is then increased linearly to \(25~\mathrm{mm}\) between \(t=0.2\) and \(1.0~\mathrm{s}\). The resulting axial loading rate is \(31.25~\mathrm{mm/s}\), corresponding to a nominal strain rate of approximately \(3.08\times10^{-2}~\mathrm{s}^{-1}\). The analysis is performed with Abaqus/Explicit with geometric nonlinearity disabled. Every GP-SSM material point uses the frozen-memory elastic unloading rule of Section~\ref{sec:case3}, since the structural loading condition produces unloading in some elements while others continue to load.

Figure~\ref{fig:case4_structural}(a) shows that the GP-SSM closely tracks the full force-displacement relation of the KCC reference, including its initial stiffness, peak load, and softening. Its peak force is only \(1.4\%\) higher and occurs about \(1.6~\mathrm{mm}\) later, indicating a slightly more ductile pre-peak response. After the peak, the GP-SSM loses load-carrying capacity slightly faster and predicts a reaction force about \(4.3\%\) lower at \(25~\mathrm{mm}\), suggesting stronger post-peak degradation or localization.
Figure~\ref{fig:case4_structural}(b) shows the element-wise axial stress--strain trajectories for both models. Before the peak, the KCC and GP-SSM trajectories nearly coincide. During post-peak softening, however, the KCC element responses (in orange) and the GP-SSP element responses differ.
The same contrast appears in the distribution of the final elemental axial strains for the two models, shown in Fig.~\ref{fig:case4_structural}(c). Both models leave most elements at the elastically unloaded value, but the KCC elements in the tail tend to fall into discrete clusters, while the GP-SSM counts form a single smooth tail with a slightly wider spread. Figure~\ref{fig:case4_structural}(d) shows that this difference in local description does not change the deformation pattern. The GP-SSM band is slightly more intensely localized, consistent with the faster post-peak load drop in Fig.~\ref{fig:case4_structural}(a).

Two mechanisms contribute to the branching observed in Figure~\ref{fig:case4_structural}(b) and (c). First, the KCC implementation regularizes softening using an element-size factor derived from the element volume relative to the prescribed localization width. Discrete element-size classes can therefore generate discrete softening slopes. Second, spatial variations in confinement and boundary restraint lead different integration points to distinct pressure histories and residual-strength states. The GP-SSM represents the response through \(\z\), \(\eeq\), the previous stress invariants \(I_1\) and \(J_2\), and the six strain components. Element size and structural location are not included explicitly because the formulation targets a local material response learned from stress--strain histories rather than a mesh-dependent representation of the KCC branches.

When multiple KCC response branches correspond to inputs that are indistinguishable within the current GP-SSM feature space, the Gaussian-likelihood GP cannot represent those branches separately; its posterior mean instead tends toward their conditional average. It consequently averages the KCC branches into a central band, as would a deterministic neural network trained with a squared-error loss. The structural force nevertheless remains accurate because it is a sequence of volume integrals of the element stresses (or energy), so branchwise deviations can balance through spatial averaging even when individual element paths are not recovered. The agreement of the force--displacement curves therefore shows that the volume-averaged response is represented accurately despite the different local descriptions of post-peak softening. Reproducing the individual KCC branches would constitute a different modeling objective and would require explicit structural descriptors or a multimodal output representation. 

These results support a simple conclusion: a latent state learned from stress--strain sequences alone, with no measured or prescribed internal variable, can be accurate and numerically stable enough to serve as the local history variable in a commercial explicit solver, from a single integration point up to a full structural response.
Nothing in this formulation is specific to strain softening, moreover. The model assumes neither that strength degrades, nor that a peak exists, nor that deformation localizes. Its only material-specific commitments are the invariant descriptors chosen to drive the latent transition. The same structure therefore applies to any path-dependent material for which stress--strain histories can be recorded, including work-hardening and cyclically loaded metals and rate-dependent polymers, with the descriptor set adapted to the mechanisms that govern each.

\section{Discussion}
\label{sec:discussion}

\subsection{Latent state as learned constitutive history}

The primary conceptual contribution of our framework is the interpretation of the state-space latent variable as a learned constitutive history variable that is sufficient to make predictions of material response. Classical inelasticity introduces internal variables to make the stress depend on the past. The present model preserves that incremental structure, but learns the history representation implicitly from observable stress--strain sequences. This avoids prescribing a particular damage or plasticity variable while still providing the finite element solver with state variables that can be updated locally.

The latent state should not be overinterpreted as a unique physical damage measure. Different latent coordinates may rotate, rescale, or mix without changing the stress prediction. Nevertheless, their monotonic evolution under loading, and their dependence on equivalent strain and stress invariants, make them useful low-dimensional summaries of irreversible loading history. The frozen-memory elastic unloading rule introduced in Case~3 (Section~\ref{sec:case3}) completes this interpretation. Because the latent state is only allowed to evolve while equivalent strain increases, and is otherwise held fixed while the material point responds elastically, the state behaves as a genuine memory variable rather than a function of the instantaneous strain alone. In this sense, the model is closer to an internal-variable constitutive update than to a memoryless stress regressor.

\subsection{Relation to recurrent neural constitutive models}

The formulation differs from conventional gated recurrent constitutive surrogates based on long-short-term-memory (LSTM) or gated recurrent unit (GRU) architectures, in which increasing the width or depth of the recurrent architecture typically increases the dimension of the hidden state propagated between loading increments \citep{mozaffar2019deep,gorji2020rnn}. Consequently, these models often employ recurrent state vectors with many hidden components whose individual coordinates generally have no prescribed mechanical interpretation. In the present formulation, the capacity of the transition network is decoupled from the dimension of the persistent material memory. The hidden-layer activations are discarded after each update, and only the two-dimensional latent state $\mathbf z_t$ is retained as the constitutive history variable.

The two latent coordinates should not be interpreted as uniquely identified physical variables. Their evolution is nevertheless more mechanically structured than an unconstrained recurrent hidden state. Equivalent-strain progression determines when and by how much the state evolves, the softplus transition enforces monotonic accumulation, and the previous-step invariants $I_1$ and $J_2$ introduce pressure and deviatoric sensitivity. The current strain tensor is retained separately in the stress map, allowing invariant quantities to compress loading history without discarding the tensorial information required for stress prediction.

The Minimal State Cell (MSC) and its self-consistent linearized extension (LMSC) approach \citep{bonatti2021oneforall,bonatti2022self} decouples network capacity from persistent-state dimension and reduces sensitivity to loading-path discretization.
In contrast, the present formulation expresses latent-state evolution through a mechanics-informed evolution law driven by invariant quantities, in which the learned transition represents an evolution rate that is scaled by the equivalent-strain increment. This construction follows the structure of classical internal-variable evolution equations more closely than a generic recurrent hidden-state update while retaining a learned latent representation. 

For this class of models, the latent dimension should be chosen as the
smallest dimension that makes the constitutive update adequately
Markovian. The number of state variables required by a conventional
model representing similar material mechanisms can provide a useful
initial reference, but does not imply a one-to-one correspondence.
Too few latent variables may fail to distinguish materially
different loading histories, whereas excessive dimensions introduce
redundant, weakly identified and tightly coupled coordinates and may increase the risk of
overfitting.

Additional constraints, such as irreversibility
and thermodynamic consistency expressed through nonnegative mechanical
dissipation, can restrict the admissible space of latent evolutions.
Related thermodynamics- and mechanics-informed architectures illustrate
how such structure can be embedded in learned constitutive models and
show improved robustness to noise and inputs outside the training domain
\citep{masi2021thermodynamics,asad2022mechanics}. Thus, the proposed approach offers numerous opportunities for growth and extensions to wide-ranging material classes with increasingly complex deformation mechanisms in the future.

The GP stress map provides an additional distinction.
Rather than using a deterministic neural network for all components of the constitutive update, the present model separates history evolution from stress prediction and employs a sparse multi-output Gaussian process to map the latent state and current strain tensor to the stress tensor. Besides its data efficiency, the GP retains an explicit posterior representation that can be used offline to assess epistemic uncertainty, identify extrapolation, and guide new data acquisition. The present VUMAT evaluates only the posterior mean because explicit finite-element analysis requires inexpensive local constitutive updates, but the probabilistic formulation remains valuable for model assessment.
What makes the posterior useful in practice is its decomposition. As shown in Section~\ref{sec:uncertainty-decomposition}, the epistemic component measures distance from the training support and is the only contribution that additional experiments can reduce, whereas the likelihood noise is a fixed pedestal common to every sequence.
Since this noise is the same for every loading path, it does not affect how paths compare, and the epistemic term alone ranks candidate experiments by how poorly the existing data support them. This ranking requires only the proposed state--strain query, so it can be computed before the experiment is performed.
The model therefore supplies its own acquisition and quality criterion. New tests can be targeted where the posterior is widest rather than by uniform coverage or engineering intuition, and the same quantity flags when a structural calculation leaves training regime.

Note that this capability is absent from the fixed-form alternative. A least-squares calibration of the KCC strength surfaces returns a single parameter set and no internal measure of where that set is unsupported. As \ref{app:kcc-calibration} shows, the calibration data are reproduced by the KCC model about equally well by a whole family of parameter sets rather than by one, so which member the fit returns is settled by the optimiser and the starting point. The resulting model does not describe this ambiguity, and reports the same unqualified confidence everywhere it is subsequently evaluated. Adding experiments does not systematically reduce this ambiguity, because the residual reflects the assumed functional form rather than a lack of data. The learned formulation instead reports where it is uninformed and improves in exactly those regions when data are supplied, which is the practical value of carrying uncertainty through the constitutive model.

\subsection{Limitations and future work}

Because this finite element deployment uses an explicit dynamic algorithm, the lateral stress response at low confinement is inherently less numerically stable, a characteristic of explicit dynamic time integration rather than of the learned constitutive update itself. Since the training histories were assembled to characterize behavior under moderate-to-high confinement, they do not specifically target this low-pressure regime, and predictive accuracy there is correspondingly expected to be weaker and warrants further investigation. In addition, the elastic unloading rule of Case~3 is a minimal, non-learned closure. It freezes the latent memory and reuses the initial elastic stiffness rather than learning an unloading/reloading stiffness directly from data, and it does not represent nonlinear unloading behavior such as material hysteresis or the stiffness softening caused by crack propagation.

Future work could further strengthen the physical interpretation of the latent variables by coupling the GP-SSM to multiscale modeling frameworks. Lower-scale simulations could provide information on mechanisms such as microcrack growth, frictional sliding, and plastic deformation, allowing the latent representation to be learned jointly from macroscopic stress--strain histories and microstructural descriptors. Additional consistency constraints could link latent evolution to these mechanisms and their contributions to the macroscopic response. This would extend the present interpretation of the latent state as accumulated material memory toward a representation informed by processes across scales.

The framework could also be extended to other material classes, including viscoelastic polymers and viscoplastic metals. For rate-dependent materials, the latent transition could incorporate physical time, strain rate, and temperature, with evolution integrated over the physical time increment. This would allow the state to evolve during sustained loading and strain holds, providing a route to representing creep, stress relaxation, and rate-sensitive yielding. The latent evolution rules could be tailored to the relevant mechanisms, combining monotonic components for irreversible accumulation with signed components for recoverable processes. Training histories spanning different loading rates, hold durations, and temperatures would support the identification and validation of these extensions.

\section{Conclusions}
\label{sec:conclusion}

This study develops a general GP-SSM framework for modeling the path-dependent mechanical response of rate-independent materials and applies the approach to finite element modeling of quasi-brittle materials. By learning latent history variables through variational inference and constraining their evolution with mechanics-informed structure, the framework constructs an incremental constitutive model using only observable stress--strain quantities. The principal features are:

\begin{enumerate}
    \item \textbf{The framework constructs a path-dependent constitutive model entirely from observable quantities.} Variational inference learns a latent representation of material history directly from stress--strain sequences, eliminating the need to prescribe or measure model-specific damage, plastic-strain, or hardening variables.

    \item \textbf{Mechanics-informed structure turns the learned state into a compact constitutive memory.} The two-dimensional state evolves through a nonnegative rate scaled by equivalent-strain progression and conditioned on the pressure and deviatoric invariants \(I_1\) and \(J_2\). This construction suppresses spurious drift, enforces cumulative evolution during loading, and decouples the dimension of the persistent material state from the capacity of the transition network. Retaining the current strain tensor in the stress map further preserves the full tensorial information needed for six-component stress prediction.

    \item \textbf{The sparse GP stress map accurately reproduces held-out sequential responses while retaining probabilistic information.} Full-sequence closed-loop rollouts captured the confinement-dependent loading, peak, softening, and residual behavior. The predictive intervals widened in the more difficult low-confinement cases, demonstrating how the GP posterior can complement point predictions with an internal indicator of data support and extrapolation risk.

    \item \textbf{The learned constitutive update can be implemented into a lightweight, self-contained Abaqus/Explicit material model.} The exported evaluator requires no machine-learning library and advances only local state variables at each integration point. Under a changing-confinement path not included in training, the VUMAT tracked the pressure increase and reproduced the response from traditional model through peak and post-peak softening. This establishes a complete route from observable material data to an incrementally evaluated finite element constitutive law.

    \item \textbf{The state-space formulation extends naturally from material-point loading to unloading and structural response.} Freezing the latent memory below the historical maximum equivalent strain preserves irreversible history while allowing stable unloading--reloading. More significantly, the structural simulation shows that a constitutive update learned solely from material-point histories transfers to a structural boundary-value problem without any training on structural force--displacement data. This scale transfer is the essential requirement for a data-driven constitutive law to be of practical use in engineering analysis, since it allows a model calibrated with element-level material tests to be applied directly to structural components.

    \item \textbf{The framework provides a modular foundation for broader data-driven constitutive modeling.} Its transition law, probabilistic stress map, and finite element deployment are separable components that can be extended independently. Further, because the formulation itself is agnostic to the specific physics of quasi-brittle damage, the same latent-state architecture extends naturally to other history-dependent material classes, including rate-independent and rate-dependent plasticity and viscoelastic or viscoplastic polymers, by retraining on the corresponding stress--strain histories and adapting the invariant conditioning to the governing physics of each material class.

    \item \textbf{Strain localization can be treated separately from the learned local constitutive response.} Localization may be controlled through a material-point or element-size-dependent regularization, or through a field-coupled nonlocal or gradient formulation. In the present structural verification, the GP-SSM captured the global force--displacement response even though the statistical distributions of element-level stress--strain trajectories may differ. The global structural response and the local trajectory ensemble are therefore complementary validation targets: the former reflects the integrated load-carrying behavior, whereas the latter depends on the particular constitutive and localization design.
\end{enumerate}

These results demonstrate the principal advantages of the proposed approach. It learns constitutive history from observable data, embeds that history in a compact and mechanically structured state, retains a probabilistic measure of model support, and produces a local update suitable for finite element calculations. The framework thereby connects state-space learning with the internal-variable structure of classical constitutive mechanics while reducing the need for prescribed phenomenological history variables.

\section*{Declaration of competing interest}

The authors declare that they have no known competing financial interests or personal relationships that could have appeared to influence the work reported in this paper.

\section*{Acknowledgement}
This work was sponsored by the Department of Defense, Defense
Threat Reduction Agency, USA under award HDTRA1-20-2-0001 to
Johns Hopkins University and was performed through the Material
Science in Extreme Environments (MSEE) University Research Alliance
(URA). The content of the information does not necessarily reflect
the position or the policy of the federal government, and no official
endorsement should be inferred.

\section*{Data availability}

Some or all data, models, or code that support the findings of this
study are available from the corresponding author upon reasonable
request.

\appendix
\section{Implementation details}
\label{sec:implementation}

The present implementation uses a two-dimensional latent state, a neural transition network with hidden width 64, 150 inducing points for the sparse GP, and two cross-validation folds. The training split is 12 sequences for training, 5 for validation, and 3 for testing. The finite element implementation stores the two latent variables, accumulated strain components, and previous-step stress invariants as state variables. These details are included in Table~\ref{tab:implementation} for reproducibility.

\begin{table}[H]
\centering
\caption{Implementation settings used in the present study.}
\label{tab:implementation}
\begin{tabular}{ll}
\toprule
Quantity & Value \\
\midrule
Number of stress--strain histories ($N_{\text{hist}}$) & 20 \\
Train/validation/test split (\(N_{\text{train}}/N_{\text{val}}/N_{\text{test}}\)) & 12/5/3 sequences \\
Cross-validation folds & 2 \\
Latent dimension (\(d_z\)) & 2 \\
Transition MLP architecture (\(f_\theta\)) & \(5\rightarrow64\rightarrow64\rightarrow64\rightarrow2\) \\
Transition activation & tanh hidden layers, softplus output \\
GP input dimension (\(d_z+6\)) & 8 \\
GP output dimension & 6 \\
Inducing points (\(M\)) & 150 \\
Maximum epochs & 5000 \\
Learning rate & \(5\times10^{-3}\) \\
Latent-rollout Monte Carlo samples in loss \(S\) & 3 \\
Latent-rollout Monte Carlo samples at evaluation \(S_{\mathrm{eval}}\) & 100 \\
Early-stopping patience & 100 evaluation intervals \\
Validation mode & autoregressive rollout \\
Best validation RMSE (normalized stress) & 0.034698 at epoch 3100 \\
Exported scalar parameters & 10994 \\
VUMAT state variables & 10 \\
Elastic initialization threshold & \(\eeq<5\times10^{-4}\) \\
Elastic constants & \(K=16370~\mathrm{MPa}\), \(G=12790~\mathrm{MPa}\) \\
\bottomrule
\end{tabular}
\end{table}

The exported model is read by the VUMAT from a plain-text parameter file. A separate diagnostic output can be written during analysis to compare Abaqus field output, including logarithmic strain, stress, and state variables, with the internal GP-SSM material-point history.

\section{Normalization statistics}
\label{sec:statistics}
The normalization statistics used for the exported model in the present verification (Section~\ref{sec:data}), following the convention \(\tilde{x}=(x-\mu_x)/s_x\), are as follows. The strain and stress means are
\begin{equation}
\boldsymbol{\mu}_{\varepsilon}= [-0.014826,\;0.013682,\;0.013682,\;0,\;0,\;0],
\end{equation}
\begin{equation}
\boldsymbol{\mu}_{\sigma}= [-52.544,\;-15.907,\;-15.907,\;0,\;0,\;0]~\mathrm{MPa},
\end{equation}
with shared component scales \(s_{\varepsilon}=0.023302\) and \(s_{\sigma}=34.0975~\mathrm{MPa}\). The scalar descriptor statistics are
\begin{equation}
\mu_{\eeq}=0.032918,
\quad s_{\eeq}=0.052752,
\quad \mu_{I_1}=-84.3490~\mathrm{MPa},
\quad s_{I_1}=75.3799~\mathrm{MPa},
\end{equation}
\begin{equation}
\mu_{J_2}=1244.2137~\mathrm{MPa}^2,
\quad s_{J_2}=2093.0376~\mathrm{MPa}^2 .
\end{equation}

\section{Non-uniqueness of strength-surface calibration}
\label{app:kcc-calibration}
\begin{figure}[htbp]
    \centering
    \includegraphics[width=\linewidth, trim=3cm 5.5cm 2cm 5cm,clip]{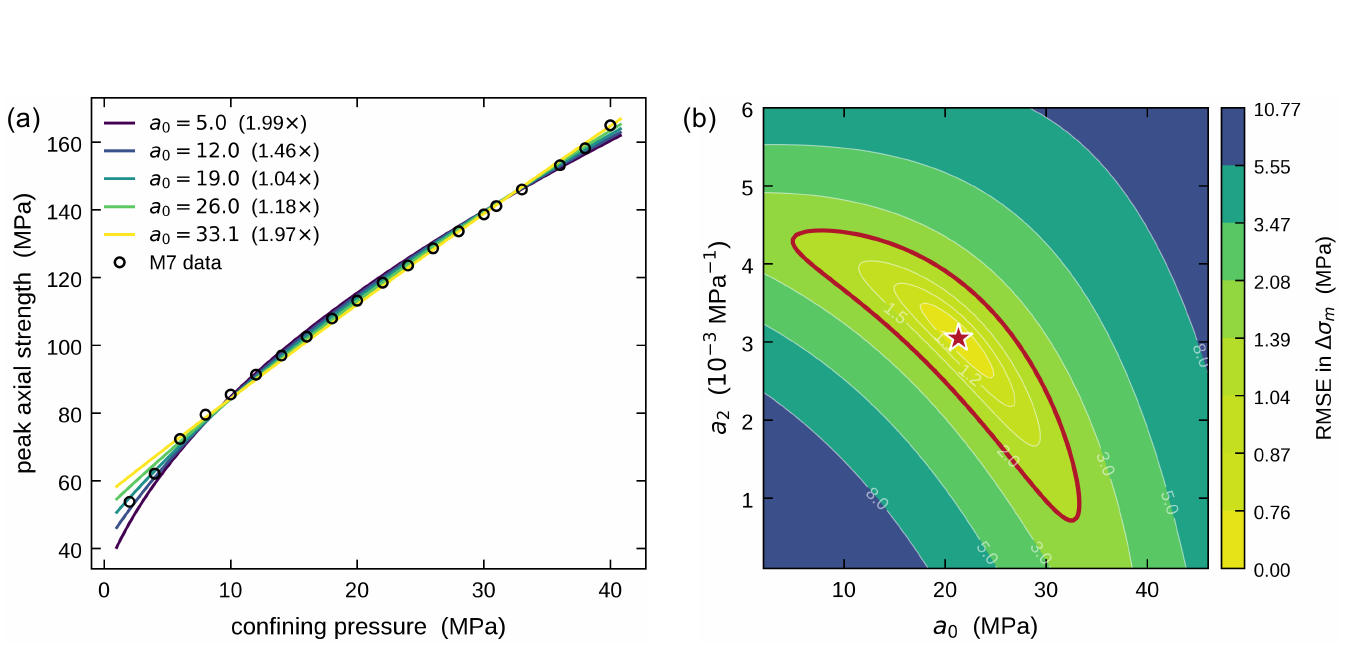}
   \caption{Non-uniqueness of the KCC maximum-surface calibration.
(a) Fits of $\Delta\sigma_m(p)$ (Eq.~\eqref{eq:kcc_surfaces}) to twenty M7
peak triaxial strengths (circles), each fixing $a_0$ at the value shown and
refitting the rest; parentheses give the RMS residual relative to the
optimum ($0.69$~MPa). Values of $a_0$ spanning nearly an order of magnitude
fit the data almost equally well, separating only below about $10$~MPa.
(b) RMS residual over the $(a_0,a_2)$ plane, with the optimum (star) and the
contour at twice its residual (red), which encloses the fits of~(a): a long
correlated valley rather than an isolated minimum, since $a_0$ and $1/a_2$
are constrained mainly through the asymptote $a_0+1/a_2$.}
    \label{fig:kcc-fitting}
\end{figure}
Section~\ref{sec:phenomenological-cost} argued that the elastic predictor is
essentially free while the boundary function must be designed from
phenomenology. This appendix makes that cost concrete for the maximum surface
$\Delta\sigma_m(p)$ of Eq.~\eqref{eq:kcc_surfaces}, fitted by nonlinear least
squares to twenty triaxial peak strengths generated with the microplane model
M7 \citep{caner2013microplane1,caner2013microplane2}, for a concrete calibrated to the same
unconfined compressive strength, elastic modulus, and Poisson's ratio as the
material of Section~\ref{sec:results} ($k_1 = 87.35\times10^{-6}$,
$E = 30448.2$~MPa, and $\nu = 0.19$).
The intent is not to argue that KCC model
cannot represent this ``new type'' of material, but that even the most favourable step of its
calibration (three parameters, a directly observable target, no time
integration) is already ill-posed.

The fit shown in Fig.~\ref{fig:kcc-fitting}(a) looks normal. The fitted surface
passes through the twenty peak strengths across the whole range of mean
pressures tested, with a root-mean-square residual of $0.69$~MPa, and no
systematic departure is visible anywhere along the curve. Judged as it would
ordinarily be judged, by how closely the calibrated surface reproduces the
data it was fitted to, the calibration is successful.
What the figure cannot show is that the parameter set producing this curve is
poorly determined. The three coefficients are strongly correlated: as
$p\to\infty$, Eq.~\eqref{eq:kcc_surfaces} approaches the finite asymptote
$a_0+1/a_2$, so $a_0$ and $1/a_2$ enter the high-pressure response only through
their sum and can be exchanged almost freely unless the data extend far enough
to constrain that asymptote, which lies well beyond the largest pressure tested
here. The least-squares objective therefore has no isolated minimum but a long
curved valley of parameter sets, shown in Fig.~\ref{fig:kcc-fitting}(b), spanning nearly an order of magnitude in
$a_0$, every one of which reproduces the data in Fig.~\ref{fig:kcc-fitting}(a)
about as closely as the curve plotted. Their predictions coincide over the
calibrated range and diverge widely outside it. Nor is this resolved by
collecting more data at the same pressures: the residual at the optimum is
systematic model-form discrepancy rather than measurement scatter, so
additional measurements force a compromise across it instead of refining the
estimate.

This is the phenomenological cost of Section~\ref{sec:phenomenological-cost}
made quantitative, and it is a property of fitting a rigid functional form to
data of limited range, not a defect specific to KCC. The GP-SSM developed in
this paper does not avoid this by having fewer parameters; indeed, it has
substantially more. What changes is that they are inferred by variational
inference directly from observed stress--strain sequences, so an additional
experiment tightens a well-posed optimisation rather than adding a further
target to a manual search over a flat, correlated objective. No surface is
designed or tabulated in advance. The posterior variance further reports where
the model is unsupported by data, which a calibrated fixed-form surface does
not provide.

\section{Training Data Generation}

Synthetic stress--strain histories were generated from single-element tests. These tests were simulated in Abaqus/Explicit using the Karagozian and Case Concrete (KCC) Release III model, implemented through a user-defined VUMAT. These simulations provide controlled reference responses for learning the dependence of stress on strain and loading history under different confinement levels.

The specimen was represented by a cube with an initial side length of \(101.6~\mathrm{mm}\), discretized using one eight-node, reduced-integration solid element (C3D8R) with enhanced hourglass control. Geometric nonlinearity was enabled. On each of the three coordinate planes \(x_i=0\), the normal velocity component was constrained to zero, while tangential motion remained unconstrained. The single integration point provided a material-point stress--strain history for each prescribed loading condition.

The material parameters were held fixed across the simulations. The density was \(2321~\mathrm{kg/m^3}\), the unconfined compressive strength was \(f_c'=45.2~\mathrm{MPa}\), and Poisson's ratio was \(\nu=0.19\). The specified bulk and shear moduli were \(K=16.37~\mathrm{GPa}\) and \(G=12.79~\mathrm{GPa}\), respectively. The localization width used by the KCC regularization was \(48~\mathrm{mm}\). The automatic parameter-generation option for normal-weight concrete supplied the remaining unspecified constitutive parameters. Strain-rate enhancement was disabled, so the reference constitutive response did not include rate-dependent strength amplification.

Each simulation followed a two-stage loading protocol. During the first stage, equal compressive pressures were applied to the three faces opposite the constrained coordinate planes and increased smoothly from zero to the target confining pressure \(p_c\) over \(0.20~\mathrm{s}\). During the second stage, these pressures were maintained, and additional axial compression was imposed by prescribing the velocity of the face initially located at \(x_1=101.6~\mathrm{mm}\). The axial velocity increased smoothly in magnitude from zero at \(t=0.20~\mathrm{s}\) to \(25.4~\mathrm{mm/s}\) at \(t=0.30~\mathrm{s}\), after which it remained constant. This velocity corresponds to a nominal axial strain-rate magnitude of \(0.25~\mathrm{s}^{-1}\), based on the initial specimen length. The second stage lasted \(0.40~\mathrm{s}\), giving a total analysis duration of \(0.60~\mathrm{s}\). Different stress-strain histories were obtained by varying the prescribed confinement \(p_c\).

Stress and strain were recorded directly within the VUMAT at successive constitutive updates, including both elastic and inelastic loading. The strain tensor was calculated from the deformation gradient as
\begin{equation}
    \mathbf{E}
    = \frac{1}{2}
    \left(\mathbf{F}^{\mathsf T}\mathbf{F}-\mathbf{I}\right),
\end{equation}
and paired with the updated Cauchy stress tensor after the constitutive correction and restoration of the hydrostatic stress contribution. Thus, the exported strains are Green--Lagrange strains, and the exported stresses are total stresses. All six independent components were retained in the order
\(11,22,33,12,23,13\), with tensorial shear strains. Although the output also contained KCC internal variables and constitutive moduli, the GP-SSM used only the strain--stress histories and descriptors calculated from them; KCC internal variables were not used as training targets.

The resulting dataset comprised 20 triaxial compression histories spanning confining pressures from \(2\) to \(40~\mathrm{MPa}\). These histories covered initial loading, peak strength, and subsequent softening toward the residual response. Complete histories were assigned to separate training, validation, and test sets. Twelve histories at \(2,6,8,12,16,20,22,24,30,33,36,\) and \(38~\mathrm{MPa}\) were used for training; five at \(10,14,18,31,\) and \(40~\mathrm{MPa}\) for validation; and three at \(4,26,\) and \(28~\mathrm{MPa}\) for testing. This partition evaluates prediction at confinement levels absent from training while preserving the sequential structure of each loading history. Subsequent preprocessing, normalization, temporal downsampling, and rotational augmentation are described in Section~\ref{sec:data}.

\bibliographystyle{elsarticle-harv}
\bibliography{references}

\end{document}